\documentclass[useAMS,usenatbib]{mn2e}
           
\usepackage[english]{babel}
\usepackage[dvips]{graphicx}
\usepackage{lscape}

\title[S0 galaxies in Fornax: masses and central stellar populations]
{The link between the masses and central stellar populations of S0 galaxies}
\author[A.G. Bedregal, A. Arag\'on-Salamanca, M.R. Merrifield and
  N. Cardiel]{A.G. Bedregal$^{1}$\thanks{E-mail: bedregal@damir.iem.csic.es},
  A. Arag\'on-Salamanca$^{1}$, M.R. Merrifield$^{1}$ and N. Cardiel$^{2}$ 
\\
$^{1}$School of Physics and Astronomy, University of Nottingham, University Park, Nottingham, NG7 2RD, UK\\
$^{2}$Departamento de Astrof\'{\i}sica, Facultad de F\'isicas, Universidad Complutense de Madrid, 28040 Madrid, Spain}
\begin{document}

\date{Accepted ***. Received ***; in original form ***}

\pagerange{\pageref{firstpage}--\pageref{lastpage}} \pubyear{2002}

\maketitle

\label{firstpage}

\begin{abstract}

Using high signal-to-noise ratio VLT/FORS2 long-slit spectroscopy,  we have
studied  the properties of the central stellar populations and dynamics of a
sample of S0 galaxies in the Fornax Cluster. The central absorption-line
indices in these galaxies correlate well with the central velocity dispersions
($\sigma_0$) in  accordance with what previous studies found for elliptical
galaxies.  However, contrary to what it is usually assumed for cluster
ellipticals,  the observed correlations seem to be driven by systematic age and
$\alpha$-element abundance variations, and not  changes in overall metallicity.
We also found that the observed  scatter in the index--$\sigma_0$ relations can
be partially explained by the rotationally-supported nature of  these systems.
Indeed, even tighter correlations exist between the line indices and the
maximum circular velocity of the galaxies.  This study suggests that the
dynamical mass is the physical property driving these correlations, and for S0
galaxies such masses have to be estimated assuming a large degree of rotational
support. The observed trends imply that the most massive  S0s have the shortest
star-formation timescales and the oldest stellar populations.

\end{abstract}


\section{Introduction}  This is the third paper of a series designed to study
S0 galaxies in the Fornax Cluster by using optical, long-slit spectroscopy
(VLT/FORS2) and archive optical and near-infrared imaging. In Bedregal et al.\
(2006a, hereafter Paper~I) we described the sample and studied the stellar
kinematics along the apparent semimajor axes of these objects. In Bedregal et
al.\ (2006b, hereafter Paper~II) we use the circular rotational velocities
obtained in Paper~I to study the Tully-Fisher relation (Tully \& Fisher 1977)
of these galaxies as part of a larger compiled sample of $\sim \, 60$ local
S0s.  In this paper we study  the sample of Fornax S0s and try to establish
links between the properties of their central stellar populations (ages,
chemical abundances) and their global properties such as  mass and dynamics. 
The ultimate goal is to find the main physical drivers  
governing the formation and evolution of S0s. 

Many previous works on stellar populations of early-type galaxies have
concentrated on the relations between velocity dispersion and Mg indices (e.g.\
Terlevich et al.\ 1981; Gorgas et al.\ 1990; Guzm\'an et al.\ 1992; Bender et
al.\ 1993; Jorgensen et al.\ 1996; Bender et al.\ 1998; Bernardi et al.\ 1998;
Colless et al.\ 1999; Jorgensen 1999; Kuntschner 2000; Poggianti et al.\ 2001;
Mehlert et al.\ 2003; Thomas et al.\ 2004), which are usually interpreted as
correlations between mass and metallicity. However, there is still controversy
about the role of relative ages and abundances of different elements in the
observed trends (Jorgensen 1999; Trager et al.\ 1998, 2000; Kuntschner et al.\
2001; Poggianti et al.\ 2001; Mehlert et al.\ 2003; Thomas et al.\ 2005) with
some recent results pointing to an important dependence of these relations on
both age and $\alpha$-element abundances (Gallazzi et al.\ 2006;
S\'anchez-Bl\'azquez et al.\ 2006).

The vast majority of these studies bundle together ellipticals and S0s 
treating them as a single population of  `early-type' galaxies. Here we
concentrate on S0s as a distinct class. Their formation history could be
very different from that of ellipticals even though some of their properties
may appear similar (see, e.g., Arag\'on-Salamanca, Bedregal and Merrifield
2006). By using a full set of 10 absorption line indices and careful comparison
with the results from stellar population synthesis models we will try to break
the degeneracies between age, overall metallicity and  $\alpha$-element
abundances.

The remainder of the paper is laid out as follows.  In
Section~\ref{sec:data:ind}, we describe the different steps followed
to obtain accurate Lick absorption line indices. In
Section~\ref{sec:ind3Avslick} we make some tests in order to compare
Lick index measurements at different spectral resolutions. In
Section~\ref{sec:modcon} we make consistency checks between the data
and the Bruzual \& Charlot (2003) models. Section~\ref{sec:results}
presents the main results and then discusses their
implications. Finally in Section~\ref{sec:conc} our conclusions are
summarised.

\section{The Data: Line Index Measurement}\label{sec:data:ind}

In this section, the main steps followed to calculate Lick/IDS
line-strength indices (Burstein et al.\ 1984, 1986; Worthey et al.\
1994; Worthey \& Ottaviani 1997) using long-slit spectra are
described. This will allow us to study scaling relations such as the
one between $\rm Mg_2$ and central velocity dispersion (e.g. Burstein
et al.\ 1988; Guzm\'{a}n et al.\ 1992; Bender, Burstein \& Faber 1993)
and to compare the results to predictions from simple stellar
population models of Bruzual \& Charlot (2003, hereafter BC03).

The basic data reduction and extraction of the kinematics are
described in Paper~I. The only change introduced at this level is a
new criterion for the binning process, following the precepts of
Cardiel et al.\ (1998). The aim of this procedure is to estimate the
minimum signal-to-noise ratio (S/N) required for the new bins in order
to obtain reasonably small uncertainties in H$\rm \beta$ line strength
measurements ($\rm \delta (H\beta)$ between 0.04 and 0.3) and so, in
the relative ages. An estimate of the minimum $\rm \delta (H\beta)$
was made for each galaxy by using different
H$\beta$-versus-metallic-index diagrams and BC03 model grids;
depending on the position of the galaxies' data points on the grids, a
given uncertainty in the indices will translate into a corresponding
error in age. In the majority of cases, an uncertainty no larger than
$\rm 4\,Gyr$ was allowed for the outermost bins. Then, the following
expression of Cardiel et al.\ was used to obtain the S/N required for
a given uncertainty $\rm \delta (H\beta)$:
\begin{equation}\label{eq:TPoffB}
\rm S/N[\AA] = \frac{7.301 - 0.2539\cdot  H\beta}{\delta (H\beta)}
\end{equation}
The S/N per $\rm \AA$ of the new bins are 100, 50 and a minimum of
$\sim$ 30 (this last value varying somewhat from galaxy to galaxy),
decreasing as the radius increases. As a result of the new binning,
the data cover median radii $\sim 2$ bulge effective radius ($R_{\rm
e}$) for this sample of S0s.

\subsection{Transformation to the Stellar Library Resolution}\label{subsec:data:lickres}
The Lick/IDS indices studied are listed in Table~\ref{ind:list}. To be
able to compare the results to simple stellar population models, the
indices must be measured at the same spectral resolution as the
model's stellar libraries.  The width of the observed spectral lines
are a convolution between the instrumental width and the velocity
dispersion of the stars projected along the line-of-sight. Therefore,
to transform the galaxies' spectra to the required resolution, they
were convolved with a Gaussian of dispersion, $\sigma$, equal to
\begin{equation}
\sigma = \sqrt{\sigma_{\rm lib}^{2} - \sigma_{\rm inst}^{2} - \sigma_{\rm gal}^{2}},\label{eq:broad}
\end{equation}
where $\sigma_{\rm lib}$ is the resolution from the stellar library,
$\sigma_{\rm inst}$ is the instrumental widening ($\sim
30\,$km$\,$s$^{-1}$, see Paper~I) and $\sigma_{\rm gal}$ is the
velocity dispersion of the galaxy at the corresponding radius, already
calculated during the extraction of the kinematics.


\begin{table}
\begin{center}
\caption{\label{ind:list}Lick indices used
  in this study (Worthey et al.\ 1994; Worthey \& Ottaviani 1997) and resolution at which each
  index was measured when Lick resolution was adopted (S\'anchez-Bl\'azquez 2004).}
\begin{tabular}{@{}l|c@{}}
\hline
 $\rm Index$ & Lick Resol.\\
         &  [$\rm km\,s^{-1}$]\\
\hline
 H$\beta$ & 225  \\
Fe5015 & 200 \\
$\rm Mg_1$ & 200 \\
$\rm Mg_2$ & 200 \\
 Mg$b$ & 200 \\
Fe5270 & 200  \\
Fe5335 & 200  \\
Fe5406 & 200  \\
Fe5709 & 200  \\
Fe5782 & 200  \\
\hline
\end{tabular}\\
\end{center}
\end{table}

 Once the spectra were broadened, line-strength indices were measured
 in the central regions of the galaxies (within $R_{\rm e}/8$ of the
 bulge) using the software {\tt INDEXF}, developed by
 one of us 
 (NC\footnote{http://www.ucm.es/info/Astrof/software/indexf/indexf.html}).
 Following the Lick/IDS index definition, the program calculates a
 \textsl{pseudocontinuum} (a local continuum level) for each spectral
 feature defined by the means within two pseudocontinuum bands located
 one at each side of the spectral feature (see Worthey et al.\ 1994
 and Worthey \& Ottaviani 1997). Then, the line index is measured with
 respect of the pseudocontinuum by integrating within the feature
 central band
\begin{equation}
\rm Index = \int_{\lambda_{c1}}^{\lambda_{c2}}\Big(1-\frac{S_{\lambda}}{C_{\lambda}}\Big) d\lambda,\label{eq:index}
\end{equation}
where $\lambda_{c1}$ and $\lambda_{c2}$ are the initial and final
wavelengths of the central band, $\rm S_{\lambda}$ is the flux of the
spectrum at a certain $\lambda$ and $\rm C_{\lambda}$ is the
corresponding pseudocontinuum flux.  This program also estimates the
uncertainties resulting from the propagation of random errors and from
the effect of uncertainties on radial velocity by performing
Monte-Carlo simulations.


The choice of BC03 models instead of other alternatives, like the
models of Thomas, Maraston \& Bender (2003, hereafter TMB03), was made
based on the higher resolution of the stellar libraries of the former
($\rm 3\,\AA$, Le Borgne et al.\ 2003), which permitted the study of
the spectral features of these galaxies in greater detail. For the
study of the main stellar population properties, however, a BC03 model
with Lick resolution had to be used (resolution from FWHM $\sim
11\,\rm \AA$ around $4000\,\rm \AA$ to $\sim 8.5\,\rm \AA$ around
$5000\,\rm \AA$) given the higher reliability of these models when
individual ages and metallicities are estimated\footnote{The Lick resolution
  models make use of the so-called {\it fitting functions} (Worthey et al.\
  1994; Worthey \& Ottaviani 1997) while the $\rm 3\,\AA$ models don't. As a
  result, the model grids of the later sometimes show irregular (unphysical)
  patters which translate in very uncertain individual ages and [Fe/H].}. We emphasise that at both,
Lick and $\rm 3\,\AA$ resolutions, we are using directly the predictions
provided from BC03. The normal distribution of BC03 models includes
predictions for both spectral resolutions, so we refer the reader to BC03 paper for
further details concerning these models. In synthesis, all
indices were measured at both resolutions by applying the previously
described procedures. Further comparisons between the two index sets
are presented in the following sections.

For NGC\,1316, NGC\,1380 and NGC\,1381, the velocity dispersions
within their central regions are higher than the $\rm 3\, \AA$ stellar
library's resolution. Therefore, the procedure described above cannot
be applied and the indices must be corrected after they have been
measured. In these cases, we proceed as follows: for each galaxy, the
best stellar-template combination for the affected bins were used as a
model of the non-convolved galaxy spectra. Using
equation~\ref{eq:broad}, these spectra were widened to the stellar
library resolution and then, in steps of $20\,$km$\,$s$^{-1}$,
convolved with Gaussians of dispersions $\sigma_{\rm gal}$ between $0$
and $400\,$km$\,$s$^{-1}$. The next step consisted of measuring all
line indices in each spectrum and calculate a correction factor for
each index, $\rm C(\sigma)_{\rm Index}$, of the form
\begin{equation}
\rm C(\sigma)_{\rm Index} = Index(0)/Index(\sigma).\label{eq:indcorfac}
\end{equation}
$\rm Index(0)$ is the index measured from the best stellar-template combination
degraded at the stellar library resolution, and $\rm Index(\sigma)$ is the index
measured from the same spectrum but widened by $\sigma
\,$km$\,$s$^{-1}$. Therefore, the corrected index is given by
\begin{equation}
\rm Index(0)=C(\sigma)_{\rm Index}\cdot Index(\sigma).\label{eq:indcor}
\end{equation}
Notice, however, that for the molecular indices $\rm Mg_1$ and $\rm Mg_2$, measured in
 magnitudes, the correction applied was of the form
\begin{equation}
\rm C(\sigma) = Index(0) - Index(\sigma).\label{eq:indcormag}
\end{equation}
The correction factors for the individual bins turned out to be
similar within each galaxy. Therefore, it was decided to apply a
single set of correction factors for each S0 by taking the average of
the individual bins' corrections and fitting the results with
polynomials of order 3.  For NGC~1316 and 1380, analogous corrections
were applied when Lick resolution was used.

The corrections were applied along the radius, $R$, to all spectral
bins for which the velocity dispersion was higher than the stellar
library's resolution ($\sigma_{{\rm gal}}^2(R) + \sigma_{{\rm inst}}^2
\ge \sigma_{{\rm lib}}^2$). For NGC~1380 and NGC~1381 the different
corrections ranged between 2--10\% of the value of the measured
indices, while for NGC~1316 they were slightly larger, typically
between 10--20\% of the original measurements. For each galaxy, an
estimate of the uncertainty in these corrections was attempted by
measuring the deviations of all the coefficients $\rm C(\sigma)_{\rm
Index}$ from each bin at a velocity dispersion of $400
\,$km$\,$s$^{-1}$. The resulting errors translate into uncertainties
of the order of $\rm 0.001\, \AA$ for the different indices.  Such
errors are negligible compared to other sources of error from the data
reduction/kinematics extraction processes, and so were neglected in
subsequent analysis.

\subsection{Emission and Lick Spectrophotometric Corrections}\label{subsec:data:emiss}
For a long time early-type galaxies were considered as gas/dust free
objects. However, subsequent work on large samples of ellipticals has
revealed that about 50\% of these objects show weak nebular emission lines in
their optical spectra (Caldwell et al. 1984; Phillips et al. 1986; Goudfrooij
et al. 1994). Measurements of different nebular lines indicates the
presence of $\rm 10^{3}$--$\rm 10^{5}\, M_{\odot}$ of ionised gas in the central regions
of these galaxies. Despite the small amount of gas, some absorption line
features can be affected by these emissions.

Weak traces of nebular emission ($\rm [O_{III}]_{\lambda 5007}$ rest frame) were found in
the spectra, which affects the measurement of the indices Fe5015 (within its
central band) and Mg$b$ (within one of the pseudocontinuum bands). To correct
these indices of emission, the best stellar-template combination was compared
to the galaxy spectrum and the spectral feature of $\rm [O_{III}]_{\lambda 5007}$ was
replaced by the corresponding section of the best stellar-template combination
before the indices were calculated. No other signatures of emission, such as
$\rm [O_{III}]_{\lambda 4959}$ or $\rm [NI]_{\lambda 5200}$, were found which could affect the
measurements of other metallic indices. 
%
%

It was also found that the index H$\beta$ has traces of contamination
from emission. The correction of this index is of particular
importance in this study given that it is the main age indicator. The
presence of emission in H$\beta$ decreases the magnitude of the
measured index and the inferred age becomes older (H$\beta$ decreases
with the age of the stellar population). Gonz\'alez (1993), studying a
sample of bright elliptical galaxies, found an empirical correlation
between the $\rm EW[O_{III}]_{\lambda 5007}$ and the EW of H$\beta$ in
emission. For his brightest galaxies he found that $\rm
\frac{H\beta_{emission}}{EW[O_{III}]_{\lambda 5007}} \sim
0.7$. However, Trager et al.\ (2000) found in a sample of 27
elliptical galaxies that this relation has variations between 0.33 and
1.25, and that the mean value is 0.6 instead of 0.7. Accordingly, it
was decided to apply the following (additive) emission correction for
the index H$\beta$:
\begin{equation}\label{eq:eps1}
\rm \Delta(H\beta) = 0.6\cdot EW[O_{III}]_{\lambda 5007},
\end{equation}
where $\rm EW[O_{III}]_{\lambda 5007}$ was measured from the residual
spectra obtained by subtracting the best combinations of stellar
templates from the galaxy spectra. The variations found by Trager et
al.\ (2000) in the ratio $\rm H\beta/[O_{III}]_{\lambda 5007}$ are
translated into uncertainties in the determination of the age of the
order of $3\%$, negligible compared to other sources of error for this
index. For the S0 galaxies analyzed here, the corrections applied to
H$\beta$ were not usually larger than 10\% of the original
measurements.

When models based on Lick/IDS libraries
(e.g. some of the models published by BC03, TMB03) are used, a
spectrophotometric correction is usually applied to the measured 
indices. This is because the Lick libraries are not calibrated in
flux, but flux-normalised by using a tungsten lamp. The indices are usually
corrected by additive factors calculated from observations 
of stars from the Lick library with the same instrumental configuration and 
photometric conditions as the data.  We lack such data in the current study;
however, there are other ways around this issue. We attempt to calibrate our
Lick resolution indices by using data from the literature. In his sample of
early-type galaxies, Kuntschner (2000) presents lower resolution, fully
calibrated central line indices for our entire sample of S0s, being an ideal
dataset for our purpose. All our indices but $\rm Fe5782$ were also measured
in his work. We rebinned our data by using a central aperture of 3.85" as in
Kuntschner (2000) in order to perform the comparison. By using the individual
measurements for each galaxy, mean corrections (offsets) were estimated for each
index. The uncertainties of these corrections, however, were quite
large. After studying the significance of the offsets by using a Student-t correlation
test, we conclude that for all the indices but $\rm Mg_1$ and $\rm Mg_2$ the
offsets were not significant at 95\% confidence level. These two molecular
indices are the most likely to be affected by spectrophotometric calibration
problems given their highly separated pseudocontinuum bands. In any case, the
corrections estimated for the remining indices were usually fare below 10\%
the value of the original measurements. Therefore, we only applied the
corrections to the $\rm Mg_1$ and $\rm Mg_2$ indices measured at Lick resolution:
\begin{equation}\label{eq:offMg1}
\rm \delta (Mg_1)_{Lick} = +0.0166 \pm 0.0056 \mbox{ mag}
\end{equation}
\begin{equation}\label{eq:offMg2}
\rm \delta (Mg_2)_{Lick} = +0.0216 \pm 0.0072 \mbox{ mag.}
\end{equation}
We stress that these corrections were applied only to Lick resolution
measurements. The stellar libraries on which $\rm 3\, \AA$ resolution models
are based have being properly calibrated, so no corrections are required for
$\rm 3\, \AA$ resolution data.

In this work, two combined indices were also used, $\rm \langle Fe\rangle$
 (Gorgas, Efstathiou \& Arag\'on-Salamanca 1990) defined as
\begin{equation}
\rm \langle Fe\rangle = \frac{Fe5270 + Fe5335}{2},\label{eq:indFeav}
\end{equation}
and [MgFe]$^{\prime}$ (Gonz\'alez 1993; Thomas, Maraston \& Bender 2003) defined as
\begin{equation}
\rm [MgFe]' = \sqrt{Mg\textsl{b}\cdot (0.72\times Fe5270 + 0.28\times Fe5335)}.\label{eq:indMgFe}
\end{equation}
[MgFe]$^{\prime}$ will be particularly important for the analysis of stellar population
parameters, like age and metallicity, given its almost null dependency on
$\alpha$-element abundance (TMB03). 
All measured and corrected central indices for the Fornax sample are presented in Tables~\ref{tab:indcent3A} (at
$\rm 3\, \AA$ resolution) and \ref{tab:indcentLick} (at Lick resolution) in
the Appendix.

 Finally, note that in different parts of this work the atomic
indices are sometimes expressed in magnitudes (following Colless et al.\ 1999)
according to the expression
\begin{equation}
\rm Index^{*}=-2.5\cdot log\Big(1-\frac{Index}{\Delta \lambda}\Big),\label{eq:indmag}
\end{equation}
were $\rm Index^{*}$ represents the index in magnitudes, $\rm Index$ is the
index expressed in $\rm \AA$, and $\Delta \lambda$ is the central bandwidth in
$\rm \AA$, listed in Table~1 of Worthey et al.\ (1994). In the particular case of $\rm \langle
Fe\rangle$, the index in magnitudes was calculated using
\begin{equation}
\rm \langle Fe\rangle^{*}=-2.5\cdot \log
\frac{1}{2}\Big(1-\frac{Fe5270}{\Delta \lambda_{Fe5270}}-\frac{Fe5335}{\Delta
  \lambda_{Fe5335}}\Big), \label{eq:indmagFeav}
\end{equation}
while [MgFe]$^{\prime}$ was always expressed in $\rm \AA$.

\section{Comparison between indices at different resolutions}\label{sec:ind3Avslick}

\begin{figure*}
\includegraphics[scale=0.6]{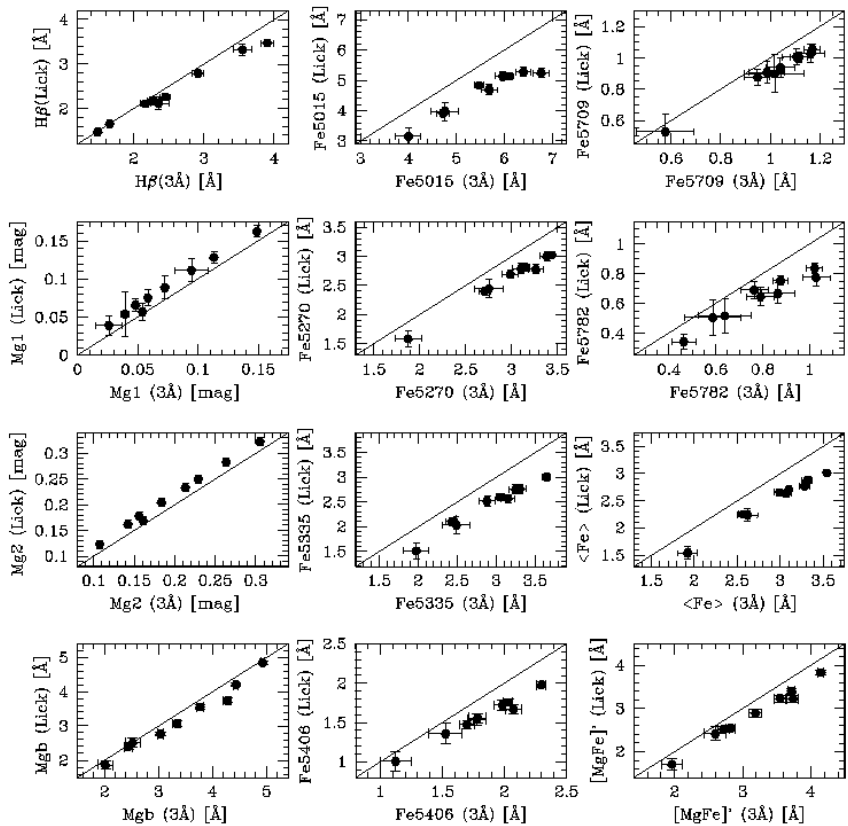}
\caption{\label{fig:ind3avsLick}\small Comparison of the central line
  indices measured at $\rm 3\, \AA$ and Lick resolutions.}
\end{figure*}

In this section, comparisons between line indices measured at different
resolutions are made in order to understand in a qualitative way the possible
effects on the ages and metallicities when inferred from degraded
spectra.
In Figure~\ref{fig:ind3avsLick}, measurements of all central indices at $\rm 3\,
\AA$ and Lick resolution are compared for the whole galaxy sample. 

\begin{figure*}
\includegraphics[scale=0.6]{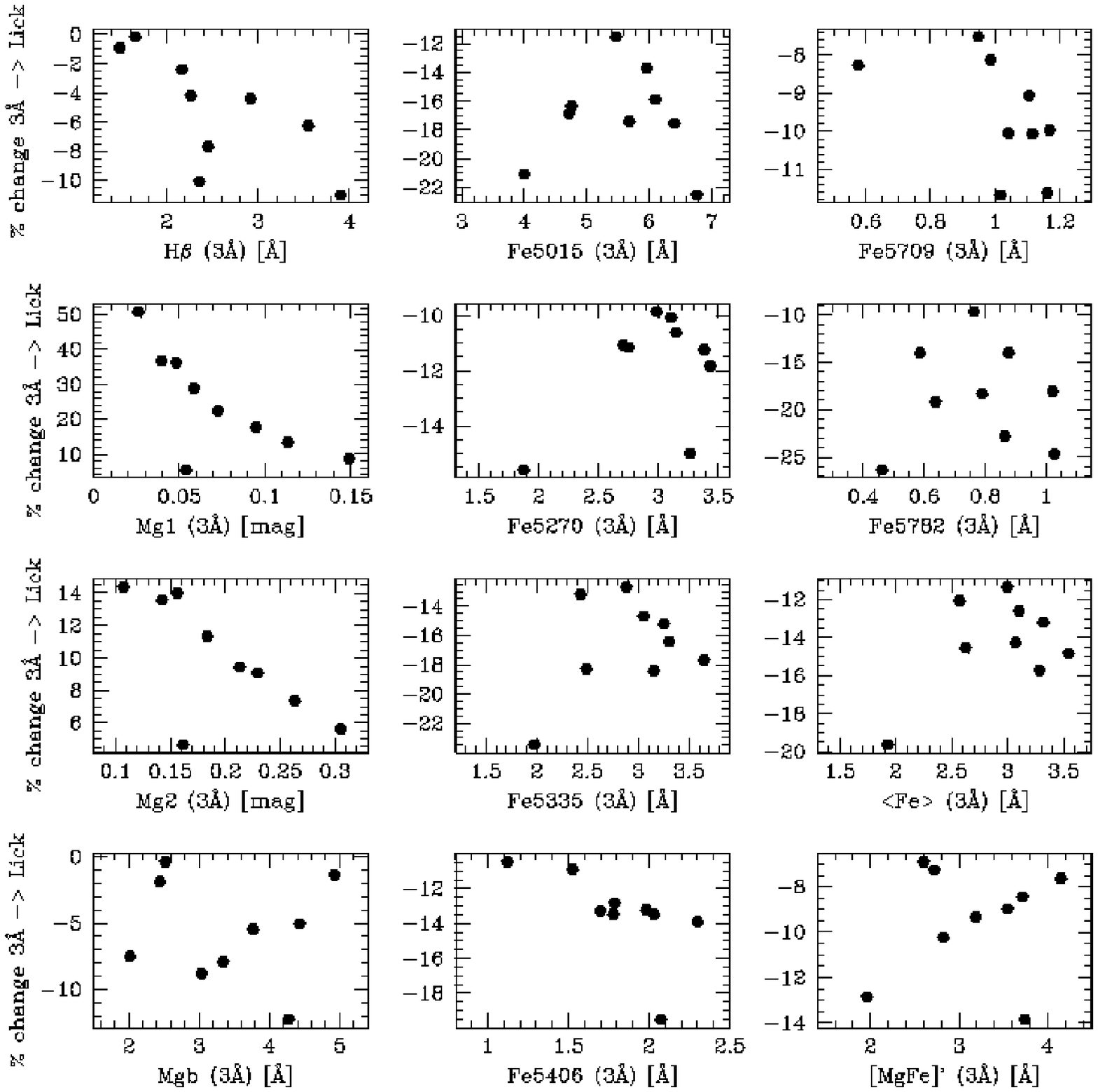}
\caption[Central line indices at $\rm 3\, \AA$ resolution versus fractional
  change (\%) when measured at the Lick
  resolution.]{\label{fig:perind3avsLick}\small Central line indices at $\rm
  3\, \AA$ resolution versus fractional change (\%) when measured at the Lick
  resolution.}
\end{figure*}

All forthcoming comparisons are made in relation to the $\rm 3\, \AA$
indices.  The H$\beta$ index suffers little degradation, where only
the galaxy NGC~1375, with the strongest H$\beta$-absorption, may
suffer from a small relative decrement of its luminosity-weighted age
estimate. The Mg indices suffer a dissimilar behaviour. While Mg$b$ presents
a very little variation between the two resolutions, the molecular
indices $\rm Mg_1$ and $\rm Mg_2$ show a systematic offset with respect to the
1:1 lines. Their larger values at Lick resolution are just a consequence of
applying the spectrophotometric correction in
Section~\ref{subsec:data:emiss}. In general, Fe indices present systematic shifts with
respect to the 1:1 lines with slightly larger differences for galaxies
with strong spectral features. The dependence on index intensity may
introduce some differences in the relative metallicities derived using
Fe features by diminishing this parameter in galaxies with strong Fe
lines. However, the relative differences in shifts among the galaxies
are not strong enough to make this an important effect. The relative
index variations between galaxies can be more easily appreciated in
Figure~\ref{fig:perind3avsLick}, where the indices at $\rm 3\, \AA$
resolution are plotted versus the fractional change when they are
measured at Lick resolution.
 
The Fe features vary, on average, by $-$15\%, while the H$\beta$ and
the Mg$b$ index variations are usually less than $-$10\%. The corresponding
panels for $\rm Mg_1$ and $\rm Mg_2$ ilustrate not only the systematic offsets
of Figure~\ref{fig:ind3avsLick} but the magnitude of the spectrophotometric
corrections applied to these two indices. The variations are particularlly
important for $\rm Mg_1$, where a median percentual change of +20\% is
found. This is the main reason why we are going to restrict the use of $\rm
Mg_1$ and $\rm Mg_2$ to the {\it qualitative} tests presented in the following
sections, without attempting to estimated individual ages and [Fe/H] with
them. In this way, Mg$b$ becomes our more reliable magnesium tracer. Concerning
the estimate of the [Mg/Fe] (or [$\alpha$/Fe]) relative abundance tracers, the fact that the Mg$b$
index presents little variation in comparison to the Fe indices will
certainly introduce changes for each individual galaxy if the
abundance tracers are measured as the differences between metallicities
(e.g. $\rm [Fe/H]_{Mgb} - [Fe/H]_{\rm \langle Fe\rangle}$). However,
the similar variations in the Fe indices experienced by all the
galaxies will result in the \textsl{relative} [Mg/Fe] abundances
between them being mostly unaffected. In any case, the index ratio
Mg$b$/$\rm \langle Fe\rangle$ provides an alternative
model-independent tracer of the $\alpha$-element abundances previously
applied by other authors (see TMB03).

In summary, in this section comparisons between line indices measured
at $\rm 3\, \AA$ and Lick resolution have been made in order to find
possible systematic differences which could affect our conclusions
when one set or the other is used. Negligible variations are found in
the H$\beta$ and Mg$b$ indices. A somewhat larger change is detected in
the Fe features, but we conclude that such effects will not introduce
important \textsl{relative} changes in the ages, metallicities and
abundance tracers between the galaxies.

\section{The Models: consistency with the data}\label{sec:modcon}

Before showing the main results, in this section some consistency
checks between the BC03 models and the data are presented. These tests
give an idea of how well the models describe the data and, in
consequence, how the model results should be interpreted.  For a
stellar population study, it is important to check if there are
significant differences in the relative spectrophotometric
calibrations of the data and the stellar libraries used by the
models. If differences are found, the results must be interpreted
accordingly. Although ages and metallicities were estimated using Lick
resolution data and models, some qualitative comparisons were also
made with the higher resolution data. Therefore, the consistency tests
were performed for both the $\rm 3\, \AA$ and Lick resolution BC03
models and data sets.

\subsection{$\rm 3\, \AA$ resolution models and data}\label{sec:modcon3A}
When using the $\rm 3\, \AA$ resolution STELIB library, it is expected
that little difference from the data will be found because both sets
of spectra were flux calibrated using spectrophotometric standard
stars.

This assumption was tested by creating index--index plots which are
almost degenerate in age and metallicity. In this way, the model grids
describe narrow bands in the index--index space, regions which should
trace the galaxies' data if the models describe accurately the
properties of these objects and if there are no problems with the
relative flux calibrations.

\begin{figure}
\begin{center}
\includegraphics[scale=0.4]{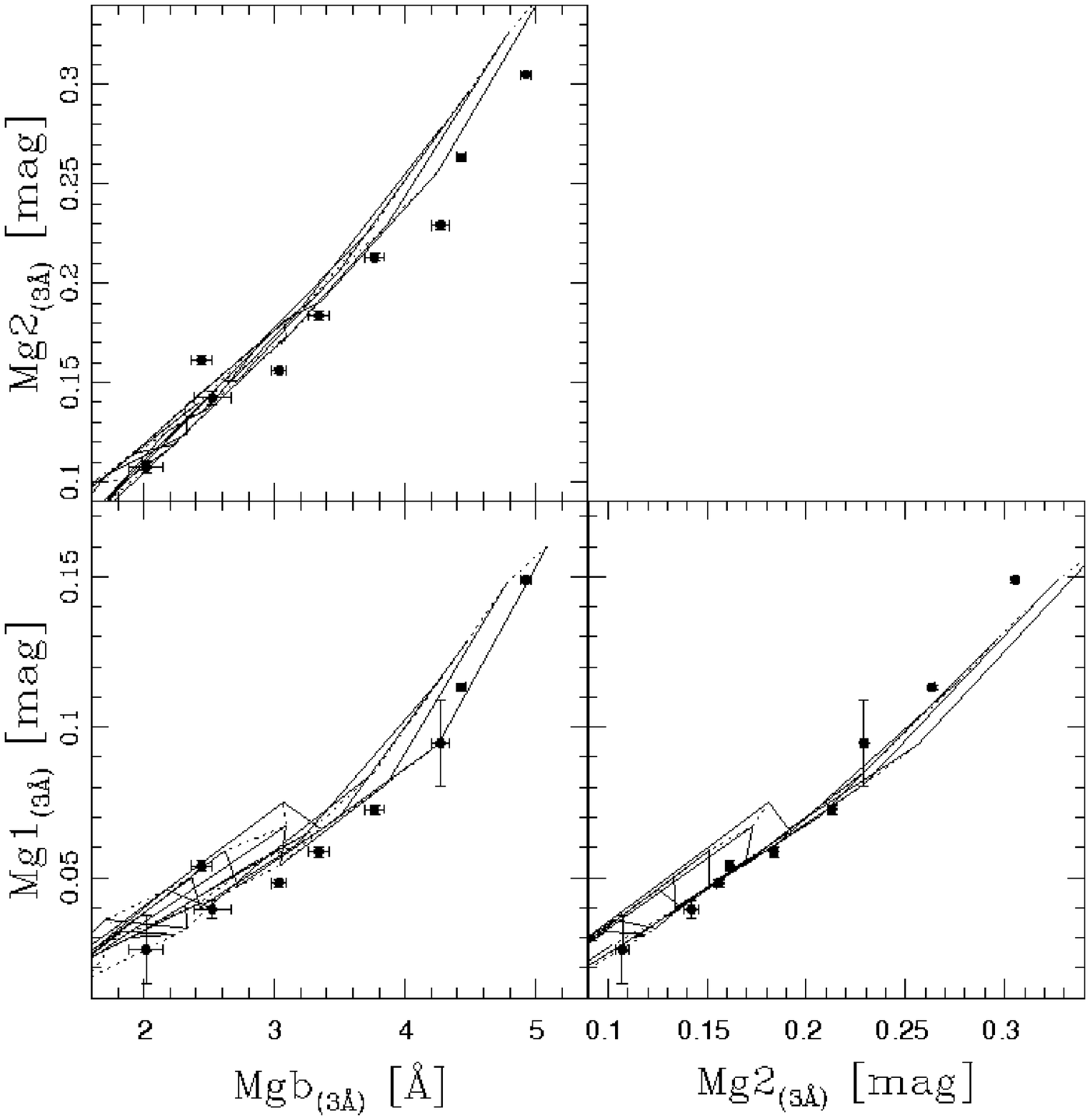}
\end{center}
\caption[Consistency test using Mg indices and BC03 models at $\rm 3\, \AA$
  resolution.]{\label{fig:Mgmodcon}\small Consistency test using Mg indices
  and BC03 models at $\rm 3\, \AA$ resolution. Folded grids are BC03 models
  and datapoints correspond to central indices (within $R_{\rm e}/8$) of S0 galaxies
  in Fornax Cluster.}
\end{figure}

\begin{figure}
\includegraphics[scale=0.4]{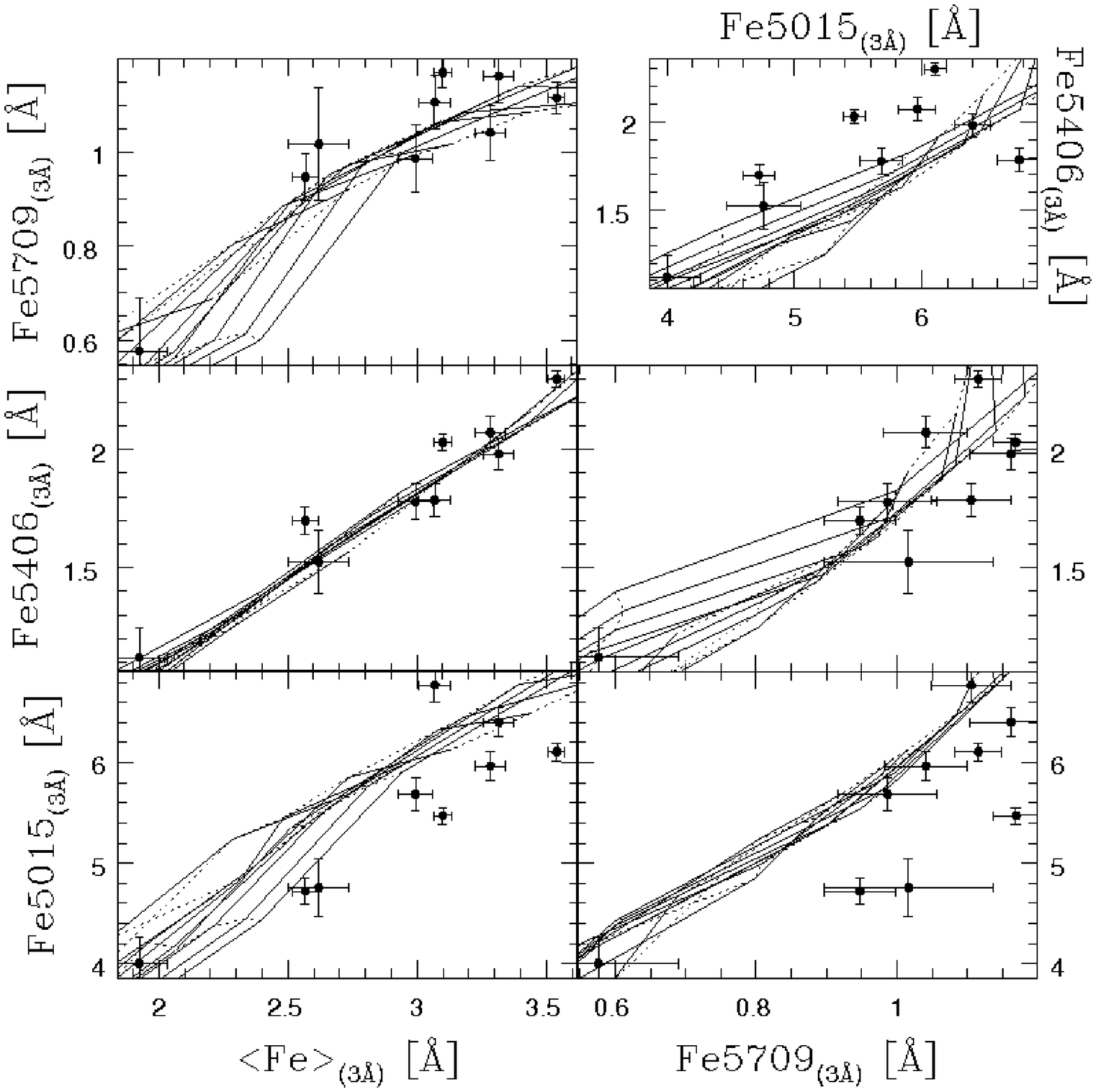}
\caption[Consistency test using Fe indices and BC03 models at $\rm 3\, \AA$
  resolution.]{\label{fig:Femodcon}\small Consistency test using Fe indices
  and BC03 models at $\rm 3\, \AA$ resolution. Folded grids are BC03 models
  and datapoints correspond to central indices (within $R_{\rm e}/8$) of S0 galaxies
  in Fornax Cluster.}
\end{figure}

In Figure~\ref{fig:Mgmodcon}, different Mg indices are plotted against
each other. The folded grids represent BC03 $\rm 3\, \AA$ resolution
models, and the line indices measured within $R_{\rm e}/8$ are plotted
for the complete S0 sample. The match between data and models is very
good, with a hint of a small systematic deviation for galaxies
with intense Mg indices (NGC~1316, 1380 and 1381). Even less pronounced, a
small offset towards lower $\rm Mg_1$ values is apparent from the $\rm Mg_1$ versus Mg$b$ panel. A similar plot is
presented in Figure~\ref{fig:Femodcon} for six Fe indices. In these
cases, the scatter from the models is larger than for the Mg indices,
but the models seem to trace the mean trends in the data. For the
three plots with Fe5015, however, there seems to be some systematic
deviation between the data and model grids. Because the central
bandpass of Fe5015 was corrected from $\rm [O_{III}]_{\lambda 5007}$
emission, this particular index must be used with caution.

Thus, as predicted, no important mismatches were found between indices
and model predictions at $\rm 3\, \AA$ resolution, suggesting that the
relative flux calibrations are consistent with each other. By assuming
solar abundances, BC03 models are able to consistently reproduce the
observations for the majority of these S0 galaxies. Therefore, in the
following sections these models can be confidently used to make
qualitative comparisons with this data set.

\subsection[Lick resolution models and data.]{Lick resolution models and data}\label{sec:modconLick}

\begin{figure}
\includegraphics[scale=0.4]{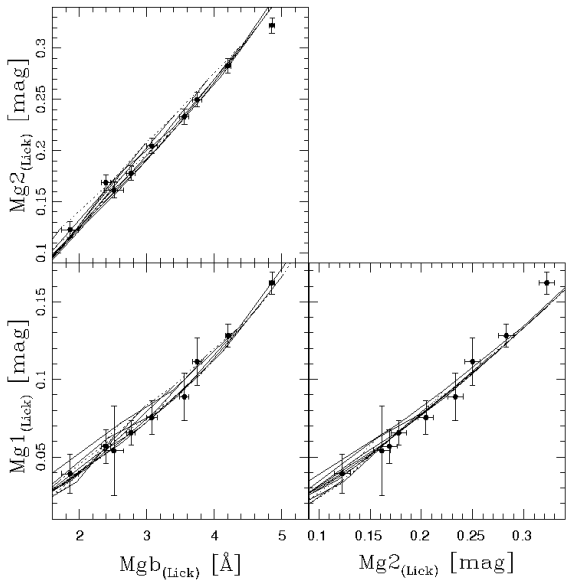}
\caption[Consistency test using Mg indices and BC03 models at the Lick
  resolution.]{\label{fig:MgmodconLick}\small Consistency test using
  Mg indices and BC03 models at the Lick resolution. Folded grids are
  BC03 models and datapoints correspond to central indices (within
  $R_{\rm e}/8$) of S0 galaxies in the Fornax Cluster.}
\end{figure}

\begin{figure}
\includegraphics[scale=0.4]{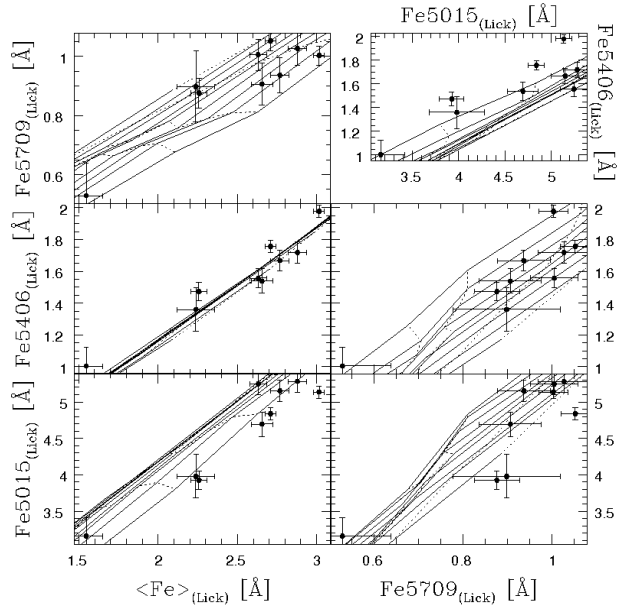}
\caption[Consistency test using Fe indices and BC03 models at the Lick
  resolution.]{\label{fig:FemodconLick}\small Consistency test using
  Fe indices and BC03 models at the Lick resolution. Folded grids are
  BC03 models and datapoints correspond to central indices (within
  $R_{\rm e}/8$) of S0 galaxies in the Fornax Cluster.}
\end{figure}

The test described in the previous section were also applied to the Lick
resolution data and models and the results are presented in Figures~\ref{fig:MgmodconLick}
and \ref{fig:FemodconLick}.

Figure~\ref{fig:FemodconLick} shows Fe indices in good agreement with
model predictions. The scatter found in the previous section and a
hint of a systematic deviation in panels involving Fe5015 are also
present here, which reinforces the idea that the Fe5015 index must be
used with caution. After the spectrophotometric correction applied in
section~\ref{subsec:data:emiss}, a very good match is also found between Mg
indices and model grids. In his study of early type galaxies in Fornax,
Kuntschner (2000, hereafter K00) found in similar tests 
discrepancies between the Mg indices and other Lick-based models
(Worthey et al.\ 1994 and Vazdekis et al.\ 1996). We do not find such
differences here. We cannot totally discard, however, the existence of such
systematic discrepancies found by K00 given ({\it i}) the large uncertainties
of our Lick spectrophotometric corrections (derived from a small sample) and
({\it ii}) the relatively large errors of our $\rm Mg_1$ indices. 

We cannot discard other sources for the individual offsets
between models and data (at both, Lick and $\rm 3\, \AA$ resolutions). For
example, the index $\rm Mg_1$ is highly dependent on the element carbon, which
could explain some deviations observed in Figures~\ref{fig:Mgmodcon} and
\ref{fig:MgmodconLick} if some peculiarities in carbon are present.

In any case, given the good agreement between models and data, and having in
mind the discussion in Section~\ref{subsec:data:lickres} (see footnote $^2$),
we can confidently use these models to calculate ages and metallicities for
individual galaxies.

\section{Results and Discussion}\label{sec:results}
In this section, the main results of the central (within $R_{\rm
e}/8$) stellar populations study are presented. As mentioned earlier,
K00 studied the central stellar populations of all ellipticals and S0s
in the Fornax Cluster using lower quality optical spectra than the ones
presented in this work. Therefore, his study will be a constant
reference in the forthcoming pages. 

From this point on, the BC03
models at $\rm 3\, \AA$ resolution were always preferred for
{\it qualitative} comparisons between the data and model grids. Also, for the
correlations between line indices versus different dynamical mass tracers (e.g.\
$\rm Index^*$--$\log(\sigma_0)$ relations) the $\rm 3\, \AA$ resolution data
were used in order to study the spectral features in grater detail. {\it Only for the
explicit calculation of ages, metallicities and abundances tracers} (e.g.\ $\rm
[Fe/H]_{Mgb} - [Fe/H]_{\rm \langle Fe\rangle}$, Mg$b$/$\rm \langle Fe\rangle$)
{\it were the models and data at the Lick resolution applied}. Unless
explicitly mentioned, the $\rm 3\, \AA$ resolution data and models
should be assumed.

\begin{figure*}
\includegraphics[scale=0.72]{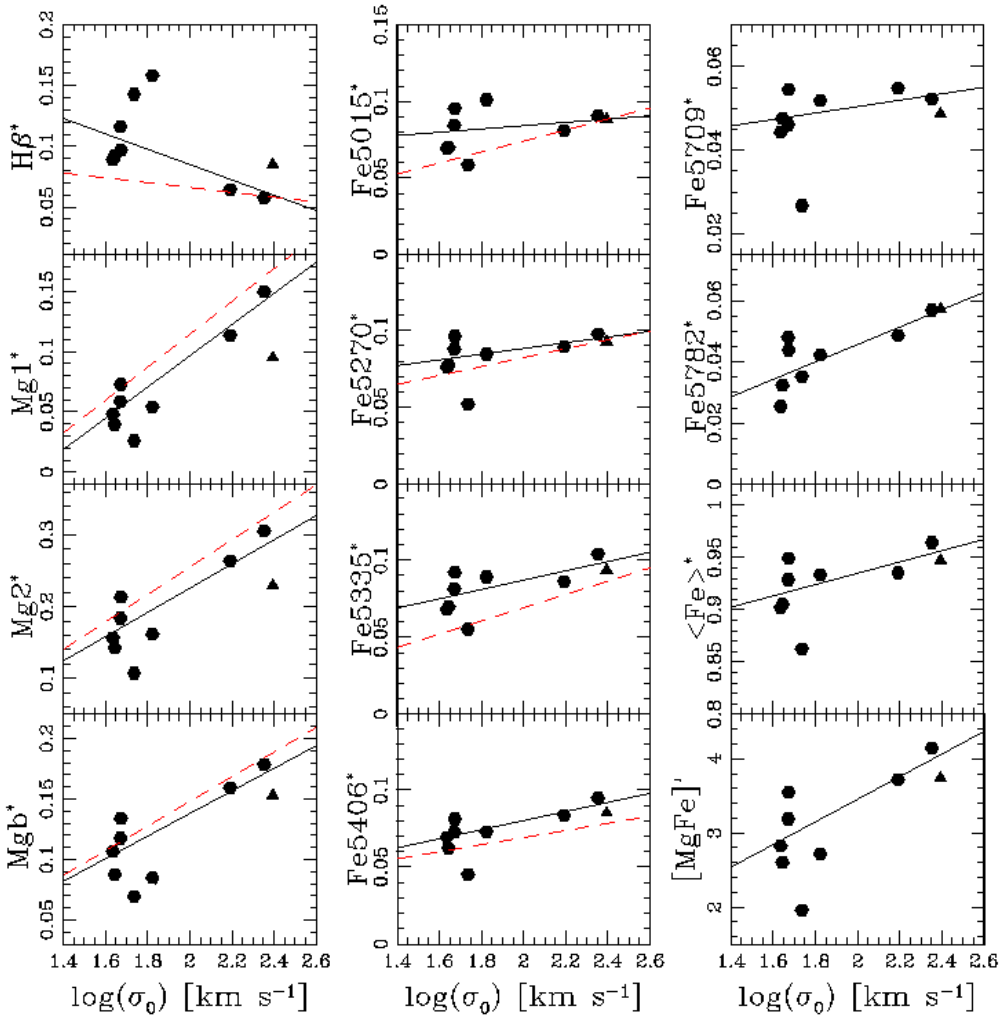}
\caption[Central line indices versus $\rm
  \log(\sigma_0)$.]{\label{fig:indsig}\small Central line indices (in
  magnitudes) versus $\rm \log(\sigma_0)$. Continuous lines are the best fits
  to the datapoints; red dashed lines represent the best fits to normal
  ellipticals in Fornax from Kuntschner (2000).}
\end{figure*}

Following previous works in this field (e.g.\ K00; Mehlert et al.\
2003; S\'anchez-Bl\'azquez et al.\ 2006), we focus our study of
central stellar populations on the relations between different line
indices and kinematical parameters of the host galaxies. The relations
between 10 line indices and the central velocity dispersion,
$\sigma_0$, are studied in order to unmask the driving variables
behind the observed slopes and scatter.  In Figure~\ref{fig:indsig},
the relations between line indices and $\rm \log(\sigma_0)$ are
presented for the sample of 9 S0s.

Rather than forming continuous trends, Figure~\ref{fig:indsig} shows
the three bright galaxies (NGC~1316, 1380 and 1381) and the rest of
faint S0s in two separate clumps, each one in opposite extremes of the
$\sigma_0$ range. However, because the sample was selected only
according to morphological classification at optical wavelengths (see
K00), there is no obvious selection bias against galaxies of
intermediate $\sigma_0$ values. Therefore, this sample will be
considered to describe the two extremes of continously-populated $\rm
Index^*$--$\log(\sigma_0)$ relations.

The plots in Figure~\ref{fig:indsig} reconfirm some of the results
already indicated by K00. First, faint S0s have a large scatter with
respect to the main relations, while the bright galaxies NGC~1380 and
NGC~1381 closely follow the relations for ellipticals estimated by K00
(red dashed lines in Figure~\ref{fig:indsig}).

Second, the merger remnant NGC~1316 (triangle in
Figure~\ref{fig:indsig}) departs significantly from the $\rm Mg_1$--
and $\rm Mg_2$--$\rm \log(\sigma_0)$ relations (at more than 4-sigma
level in both indices) towards low Mg absorption at a given $\rm
\sigma_0$, while closely following the remaining Fe--$\rm
\log(\sigma_0)$ relations. These facts, and the comparatively high
H$\beta$ absorption in the H$\beta$--$\rm \log(\sigma_0)$ diagram have
been interpreted by K00 as the signature of the presence of a younger
stellar population in this galaxy, probably formed during the
merger. Such a young population could reduce the intensity of some
metallic features by increasing the overall continuum level. This
effect would be particularly important for strong metallic features,
like the Mg triplet (indices Mg$b$ and $\rm Mg_2$) and for the
molecular indices $\rm Mg_1$ and $\rm Mg_2$ whose pseudocontinuum
level is strongly dependent on large-scale changes in the
continuum. The weaker Fe features in this spectral range could be less
influenced by this effect, explaining why NGC~1316 appears in good
agreement with the mean Fe relations.

Finally, the slopes of the Mg--$\rm \log(\sigma_0)$ trends are steeper
than the ones for the Fe lines (see fit equations in
Table~\ref{tab:indsigfit}), as K00 also found for his fits to normal
ellipticals.

\begin{table}  
 \centering
 \caption[Parameters of the linear fits ${\rm Index^*} = a + b\cdot
 \log(\sigma_0)$ of the S0 galaxies in Fornax.]{Parameters of the linear fits
 ${\rm Index^*} = a + b\cdot \log(\sigma_0)$ of the S0 galaxies in
 Fornax. $\sigma_{\rm std}$ is the standard deviation about the
 fit. \label{tab:indsigfit}} 
 \begin{tabular}{@{}lccc@{}}
  \hline
   Index  & $b\pm \delta(b)$ & $a\pm \delta(a)$ & $\sigma_{\rm std}$   \\
  \hline
  $\rm H\beta ^*$                &          $-$0.0633 $\pm$ 0.0031  &\phantom{1} 0.2114 $\pm$ 0.0066  & 0.0360 \\
  $\rm Fe5015 ^*$                &\phantom{1}  0.0107 $\pm$ 0.0021  &\phantom{1} 0.0625 $\pm$ 0.0045  & 0.0219 \\
  $\rm Mg_1 ^*$                   &\phantom{1}  0.1290 $\pm$ 0.0016  &         $-$0.1618 $\pm$ 0.0034  & 0.0229 \\
  $\rm Mg_2 ^*$                   &\phantom{1}  0.1683 $\pm$ 0.0018  &         $-$0.1111 $\pm$ 0.0038  & 0.0462 \\
  ${\rm Mg}b ^*$                 &\phantom{1}  0.0932 $\pm$ 0.0027  &         $-$0.0486 $\pm$ 0.0056  & 0.0258 \\
  $\rm Fe5270 ^*$                &\phantom{1}  0.0185 $\pm$ 0.0022  &\phantom{1} 0.0509 $\pm$ 0.0045  & 0.0161 \\
  $\rm Fe5335 ^*$                &\phantom{1}  0.0303 $\pm$ 0.0025  &\phantom{1} 0.0262 $\pm$ 0.0052  & 0.0135 \\
  $\rm Fe5406 ^*$                &\phantom{1}  0.0293 $\pm$ 0.0027  &\phantom{1} 0.0214 $\pm$ 0.0056  & 0.0112 \\
  $\rm Fe5709 ^*$                &\phantom{1}  0.0077 $\pm$ 0.0025  &\phantom{1} 0.0349 $\pm$ 0.0053  & 0.0125 \\
  $\rm Fe5782 ^*$                &\phantom{1}  0.0284 $\pm$ 0.0032  &         $-$0.0113 $\pm$ 0.0068  & 0.0071 \\
$\rm \langle Fe\rangle ^*$       &\phantom{1}  0.0538 $\pm$ 0.0037  &\phantom{1} 0.8272 $\pm$ 0.0077  & 0.0310 \\
$\rm [MgFe]'$                    &\phantom{1}  1.5212 $\pm$ 0.0801  &\phantom{1} 0.4087 $\pm$ 0.1661  & 0.5311 \\
\hline
\end{tabular}
\end{table}

\subsection{The Slopes of the Index$^*$--log($\sigma_0$) relations}\label{sec:slpIndSig}

To study which parameters are driving the observed slopes in the $\rm
Index^*$--$\log(\sigma_0)$ relations, the models of BC03 were used to
parametrise the individual indices as a function of metallicity ([Fe/H]) and $\rm
\log(age)$ such that
\begin{equation}\label{eq:indparam}
{\rm Index^*_{\rm BC03}} = A + B\cdot {\rm[Fe/H]} + C\cdot \rm{\log(age[yr])},
\end{equation}
where $\rm Index^*_{\rm BC03}$ are index values from BC03 models and
\begin{equation}\label{eq:indparamslope1}
 B = \rm \frac{\partial ({\rm Index^*_{\rm BC03}})}{\partial [Fe/H]},
\end{equation}
\begin{equation}\label{eq:indparamslope2}
 C = \rm \frac{\partial ({\rm Index^*_{\rm BC03}})}{\partial (\log(age))}.
\end{equation}
Because the galaxies in this sample cover a wide range in ages and
[Fe/H], we use models with ages from $\rm 1.0$ to $\rm 12.6\,Gyr$ and
metallicities from $\rm -0.64$ to $\rm 0.55\,dex$. The resulting
parametrisations are presented in Table~\ref{tab:paramind}. The last
two columns give the required dependence of $\rm \log(\sigma_0)$ on
[Fe/H] and age respectively, if that is the \textsl{only} driving
variable for the observed slopes in the $\rm
Index^*$--$\log(\sigma_0)$ relations. This dependency was estimated
for each index by combining the parametrisations from the models
(eqs.\ \ref{eq:indparamslope1} and \ref{eq:indparamslope2}) with the
slopes measured in Figure~\ref{fig:indsig} ($b$ in
Table~\ref{tab:indsigfit}). If [Fe/H] is considered as the driver of
the observed $\rm Index^*$--$\log(\sigma_0)$ trends

\begin{equation}\label{eq:indmet}
\frac{b}{B}=\frac{\partial ({\rm Index^*})}{\partial (\log(\sigma_0))}\cdot
\frac{\partial {\rm [Fe/H]}}{\partial ({\rm Index^*_{\rm BC03}})}=\frac{\partial {\rm [Fe/H]}}{\partial (\log(\sigma_0))}.
\end{equation}
Alternatively, if age is taken as driving variable
\begin{equation}\label{eq:indage}
\frac{b}{C}=\frac{\partial ({\rm Index^*})}{\partial (\log(\sigma_0))}\cdot
\frac{\partial (\log({\rm age}))}{\partial ({\rm Index^*_{\rm BC03}})}=\frac{\partial (\log({\rm age}))}{\partial (\log(\sigma_0))}.
\end{equation}

 Interestingly, these parametrisations show that the
 Mg--$\log(\sigma_0)$ relations depend as strongly on age as on
 metallicity. On the other hand, the slopes of the Fe indices versus
 $\log(\sigma_0)$ relations can be reproduced with weaker dependencies
 on metallicity than on age, but the differences between both
 dependencies are not very large. Naively, a much lower dependence on
 metallicity relative to the one on age would be expected for these
 metallicity tracers. Only the behaviour of the
 H$\beta^*$--$\log(\sigma_0)$ relation matches the expectations,
 presenting a slope much easier to reproduce by a small age dependence than
 by a huge metallicity one.

It is also interesting to note that not all the [Fe/H]-sensitive
indices have the same dependence on [Fe/H]; the three Mg indices ($\rm
Mg_1$, $\rm Mg_2$ and Mg$b$) required a stronger [Fe/H] dependence
than the Fe indices in order to explain their slopes in the $\rm
Index^*$--$\log(\sigma_0)$ diagrams. This would not be expected if
metallicity is the only parameter varying with $\log(\sigma_0)$.
 
Therefore, these results suggest that age could be an important driver
of the observed correlations while the role of metallicity is more
uncertain. Also a third variable, such as the relative Mg abundances,
may affect the relations by producing relative differences between
indices, each one with different sensitivities to changes in the
chemical composition of these systems (Vazdekis et al.\ 2001;
S\'anchez-Bl\'azquez et al.\ 2006).

\begin{table*}  
 \centering
 \caption[Parametrisation of the line indices using BC03 models.]{Parametrisation
 of the line indices using BC03 models: ${\rm Index^*_{\rm BC03}} = A + B\cdot \rm{[Fe/H]} +
 C\cdot \rm{\log(age[yr])}$. Last two columns: dependence of $\rm
 \log(\sigma_0)$ on metallicity (age) if this is the \textsl{only} driving variable for
 the observed slopes in the $\rm Index^*$--$\log(\sigma_0)$ relations. \label{tab:paramind}} 
 \begin{tabular}{@{}lccccc@{}}
  \hline
  Index  & $ A $ & $B\pm \delta(B)$ & $C\pm \delta(C)$ & $\rm \frac{\partial
  [Fe/H]}{\partial (\log(\sigma_0))}$& $\rm \frac{\partial (\log(age))}{\partial (\log(\sigma_0))}$   \\
  \hline
  $\rm H\beta ^*$   &\phantom{1} 0.7778  &            $-$0.0157  $\pm$ 0.0014  &           $-$0.0712  $\pm$ 0.0020    & 4.029   & 0.888  \\
  $\rm Fe5015 ^*$   &           $-$0.0854  &\phantom{1}  0.0323  $\pm$ 0.0007  &\phantom{1} 0.0177  $\pm$ 0.0011    & 0.330   & 0.601  \\
  $\rm Mg_1 ^*$      &           $-$0.4602  &\phantom{1}  0.0502  $\pm$ 0.0023  &\phantom{1} 0.0547  $\pm$ 0.0032    & 2.568   & 2.357  \\
  $\rm Mg_2 ^*$      &           $-$0.8310  &\phantom{1}  0.1075  $\pm$ 0.0031  &\phantom{1} 0.1053  $\pm$ 0.0044    & 1.564   & 1.597  \\
  ${\rm Mg}b ^*$    &           $-$0.4431  &\phantom{1}  0.0521  $\pm$ 0.0014  &\phantom{1} 0.0572  $\pm$ 0.0019    & 1.790   & 1.629  \\
  $\rm Fe5270 ^*$   &           $-$0.1509  &\phantom{1}  0.0366  $\pm$ 0.0005  &\phantom{1} 0.0244  $\pm$ 0.0007    & 0.505   & 0.757  \\
  $\rm Fe5335 ^*$   &           $-$0.1665  &\phantom{1}  0.0388  $\pm$ 0.0006  &\phantom{1} 0.0256  $\pm$ 0.0008    & 0.779   & 1.180  \\
  $\rm Fe5406 ^*$   &           $-$0.2017  &\phantom{1}  0.0419  $\pm$ 0.0006  &\phantom{1} 0.0284  $\pm$ 0.0008    & 0.699   & 1.029  \\
  $\rm Fe5709 ^*$   &           $-$0.0511  &\phantom{1}  0.0199  $\pm$ 0.0007  &\phantom{1} 0.0098  $\pm$ 0.0009    & 0.386   & 0.781  \\
  $\rm Fe5782 ^*$   &           $-$0.0856  &\phantom{1}  0.0293  $\pm$ 0.0006  &\phantom{1} 0.0137  $\pm$ 0.0009    & 0.968   & 2.074  \\ 
  \hline
 \end{tabular}
\end{table*}

\subsubsection{The [Fe/H]-- and log(age)--log($\sigma_0$) relations}
\begin{figure}
\begin{center}
\includegraphics[scale=0.43]{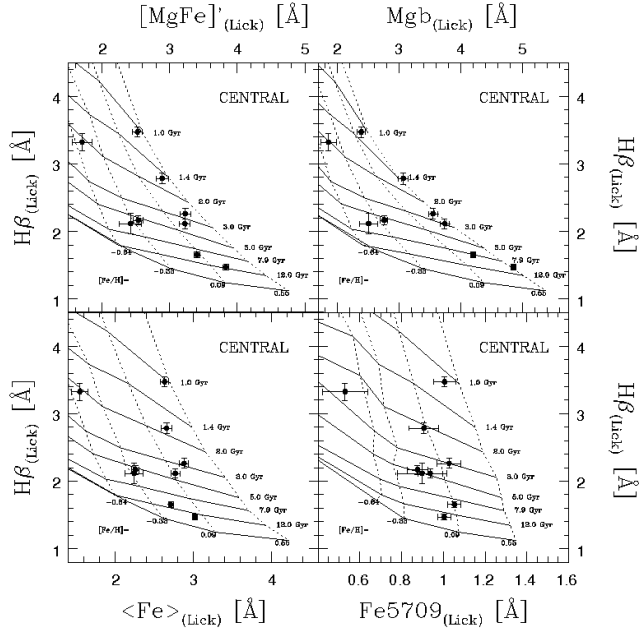}
\end{center}
\caption[Central metallic indices versus H$\beta$ for S0s in
  Fornax.]{\label{fig:grids}\small Central metallic indices versus H$\beta$
  for S0s in Fornax Cluster. The grids represent BC03 models labeled
  in terms of age and [Fe/H].}
\end{figure}

A closer comparison between the data and the BC03 models could shed
more light on the origin of the observed correlations. Individual ages
and [Fe/H] were calculated by plotting different [Fe/H]-sensitive
indices versus H$\beta$; {\it data and models at the Lick resolution were
used in this step}. The details of the procedure applied here are
described in Cardiel et al.\ (2003). Briefly, the ages and
metallicities were logarithmically-interpolated from the model grid
values by fitting the four closest cells to a datapoint with quadratic
polynomials. The procedure also considers the effect of covariance
when estimating the uncertainties for both parameters. In this way the
age-[Fe/H] degeneracy is broken and luminosity-weighted estimates of
both parameters were obtained. As an example, in
Figure~\ref{fig:grids} four of these plots are presented for the
central index values, also including the BC03 model grids labeled in
terms of age and [Fe/H]. Central ages and metallicities for different
indices are presented in Table~\ref{tab:agezcent} for all the Fornax
sample.

\begin{figure}
\begin{center}
\includegraphics[scale=0.43]{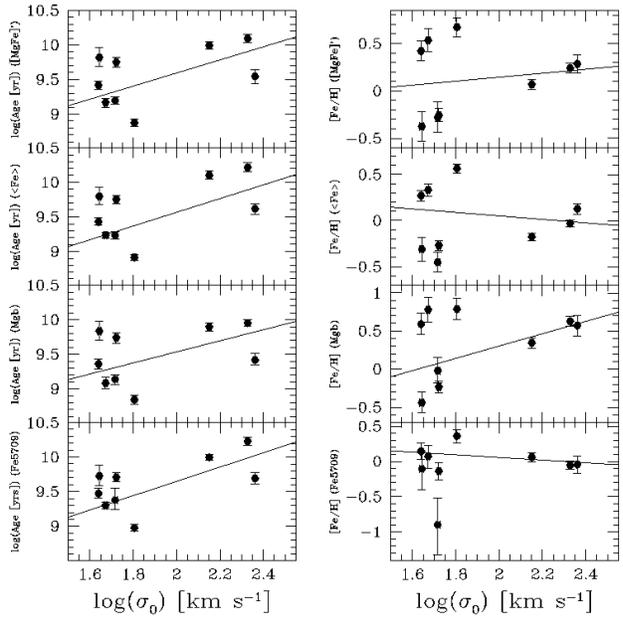}
\end{center}
\caption{\label{fig:ageZvsSigma}\small Central Ages and metallicities versus
  $\log(\sigma_0)$ for Fornax S0s. Lines represent the best linear fits to the data.}
\end{figure}

In Figure~\ref{fig:ageZvsSigma}, $\log(\sigma_0)$ is plotted against
age and [Fe/H] using different [Fe/H] sensitive indices. The
continuous lines represent the best linear fits whose slopes are
presented in Table~\ref{tab:AgeMetslope}. It is surprising that no
obvious trends are observed in any of the [Fe/H]--$\log(\sigma_0)$
plots, considering that the Index--$\log(\sigma_0)$ relations are
usually attributed, at least for ellipticals, to a chemical
composition change with galaxy mass. The slope for the
[Fe/H]--$\log(\sigma_0)$ relation using Mg$b$ seems to be an artifact
of the linear fit instead of pointing to a real correlation. Comparing
the slopes for metallicity presented in Table~\ref{tab:AgeMetslope}
with the required metallicity dependences for each
Index$^*$--$\log(\sigma_0)$ relation in Table~\ref{tab:paramind} it is
clear that [Fe/H] cannot be the main driver of any of the Index$^*$
correlations with $\log(\sigma_0)$. The plots against age present
stronger trends in the same direction as the $\rm
Index^*$--$\log(\sigma_0)$ diagrams: the three high-dispersion
galaxies seem to be, on average, older than the rest of the
sample. This result holds for all the age estimations, including the
one using [MgFe]' which is insensitive to $\alpha$-element relative
abundance variations. Comparing the slopes of these relations with the
required age dependences for the observed Index$^*$--$\log(\sigma_0)$
trends, we found that the H$\beta^*$ and almost all the Fe trends can
be explained by an age effect. However, differences in relative age do
not seem to explain the steeper slopes found for the Mg indices in
comparison to Fe, reinforcing the idea that $\alpha$-element relative
abundances are playing an important role too.  

\begin{table}  
 \centering
 \caption{Slopes from linear parametrisation of log(age) or [Fe/H] versus $\log(\sigma_0)$ for S0 galaxies in Fornax (Figure~\ref{fig:ageZvsSigma}) using different line indices as metallicity tracers. \label{tab:AgeMetslope} } 
 \begin{tabular}{@{}ll|ll@{}}
  \hline
  Parameter  & Slope & Parameter  & Slope    \\
  \hline
%
  $\log({\rm age})_{\rm [MgFe]'}$             &0.946$\pm$0.079 &${\rm [Fe/H]}_{\rm [MgFe]'}$                &\phantom{1} 0.208$\pm$0.090 \\
  $\log({\rm age})_{\rm \langle Fe \rangle}$  &0.994$\pm$0.075 &$\rm [Fe/H]_{\rm \langle Fe \rangle}$       &           $-$0.189$\pm$0.057 \\
  $\log({\rm age})_{\rm Mgb}$                 &0.803$\pm$0.080 &${\rm [Fe/H]}_{\rm Mgb}$                    &\phantom{1} 0.806$\pm$0.106 \\
  $\log({\rm age})_{\rm Fe5709}$              &1.033$\pm$0.074 &$\rm  [Fe/H]_{\rm Fe5709}$                  &           $-$0.189$\pm$0.113 \\
\hline
\end{tabular}
\end{table}

\subsubsection{The [Mg/Fe]--log($\sigma_0$) relations}

\begin{figure*}
\begin{center}
\includegraphics[scale=0.72]{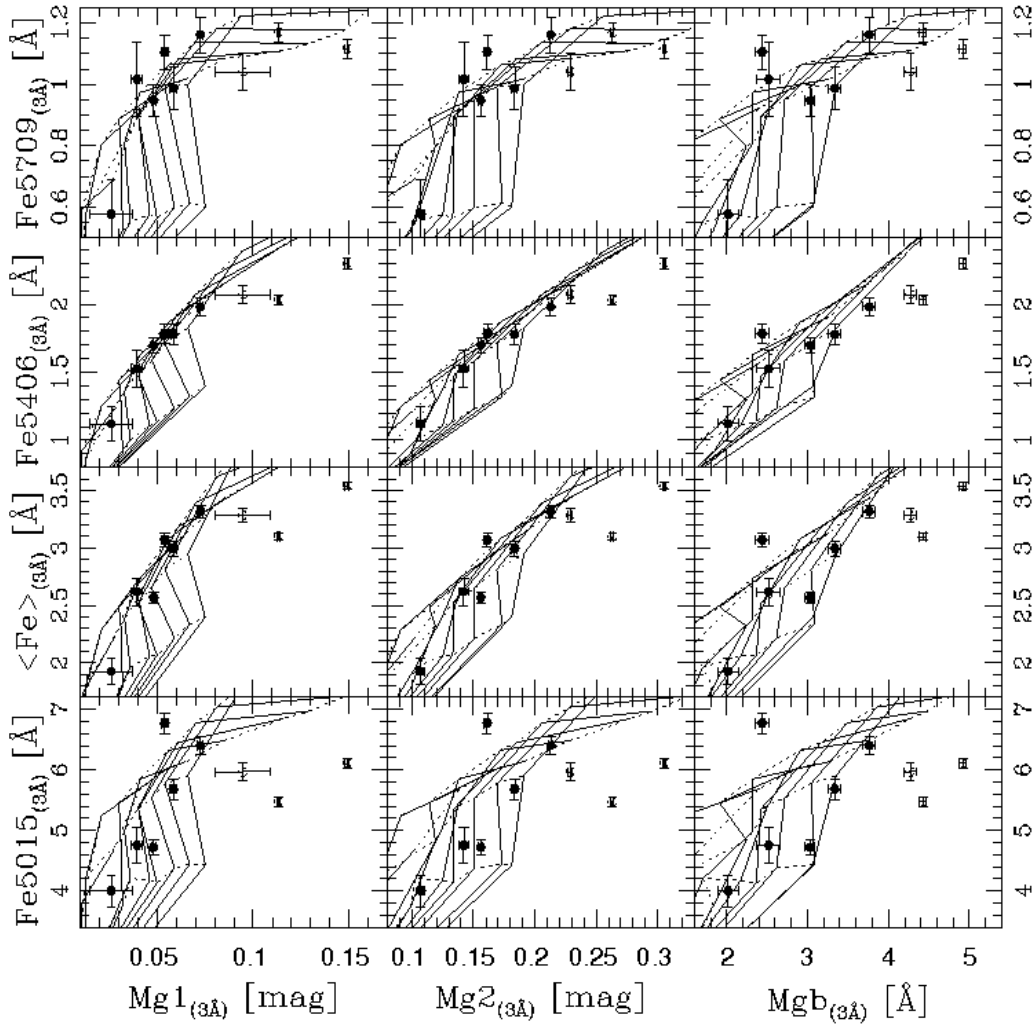}
\end{center}
\caption[Central Mg versus Fe indices for Fornax
  S0s.]{\label{fig:FevsMg}\small Central Mg versus Fe indices for Fornax
  S0s. Grids correspond to BC03 models at $\rm 3\, \AA$ resolution.}
\end{figure*}

At the beginning of this section, the possibility of a Mg (or
$\alpha$-element) abundance dependence of the $\rm
Index^*$--$\log(\sigma_0)$ relations was suggested, and this
possibility was re-enforced by the results presented above. This
alternative was further explored by plotting Mg-sensitive versus
Fe-sensitive indices and comparing the data to the solar-abundance
predictions of the models (see Worthey, Faber \& Gonzalez 1992). In
Figure~\ref{fig:FevsMg}, twelve panels plotting central Mg versus Fe
indices are presented, together with BC03 model grids. What is clear
from almost all these panels is that the three highest-dispersion
galaxies in the sample (open symbols in Figure~\ref{fig:FevsMg}) are
not accurately described by models that assume solar abundance ratios,
presenting stronger Mg indices for a given Fe index than the model
predictions. Among these three galaxies, the effect seems to be
stronger for NGC~1380 and 1381, while NGC~1316 (the merger remnant) is
usually closer to the solar-model grids. In similar plots, K00 found
that only the nucleus of NGC~1380 presented this behaviour, suggesting
that this galaxy is Mg overabundant compared to Fe.

In order to trace the relative Mg overabundances we
compare the relative metallicities derived from Fe- and Mg-sensitive
indices and the same age indicator (H$\beta$ in this case). At this point, it
is worthy to stress that we are not attempting to derive absolute Mg (or
$\alpha$-elements) abundances. By comparing different metallicity estimates
(and by using Mg$b$/$\rm \langle Fe\rangle$, see below) we aim to {\it trace}
the {\it relative} Mg abundances. All our results and conclusions are
independent of absolute abundance ratios.

\begin{figure}
\begin{center}
\includegraphics[scale=0.38]{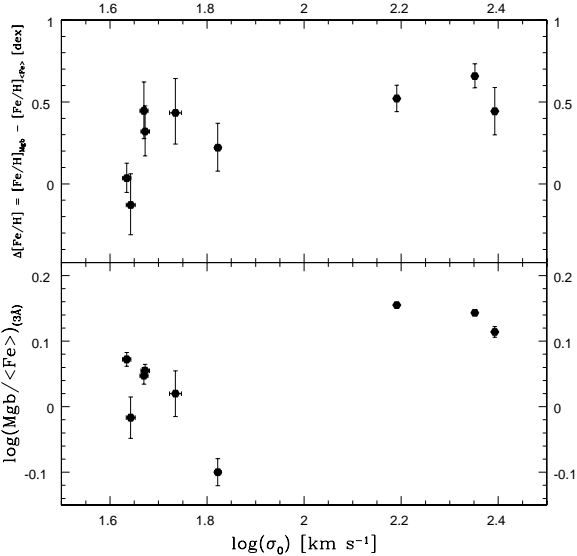}
\end{center}
\caption{\label{fig:abund}\small Mg/Fe relative abundance tracers versus $\log(\sigma_0)$ for Fornax
  S0s.}
\end{figure}

 In the top
panel of Figure~\ref{fig:abund}, $\log(\sigma_0)$ is plotted against
the difference of the metallicities calculated by using Mg$b$ and $\rm
\langle Fe\rangle$ as metallicity indicators. Data and models at the
Lick resolution were used to estimate these metallicities. It is
apparent from this plot that bright S0s are, on average, Mg
overabundant relative to fainter galaxies; however the uncertainties
coming from the models make this difference only marginally
significant. The bottom panel of Figure~\ref{fig:abund} shows the
index quotient Mg$b$/$\rm \langle Fe\rangle$ versus the velocity
dispersion. The ratio of these two indices has been used by TMB03 to
calibrate the [$\alpha$/Fe] abundance ratio of their models, so it
traces the relative behaviour of the Mg
overabundance and, equally important, it provides a model-independent test. With
much reduced errors, it is clear that bright S0s have larger
Mg$b$/$\rm \langle Fe\rangle$ ratios than fainter objects. This
model-independent test can explain why the slopes of the $\rm
Index^*$--$\log(\sigma_0)$ correlations are steeper for the Mg indices
than for the Fe ones, confirming the previous results of this
section. The overall results also imply the existence of a correlation
between age and Mg$b$/$\rm \langle Fe\rangle$ (or [$\alpha$/Fe]; see
Figure~\ref{fig:AgeZAbundvseachother}, bottom panel in first column), in
agreement to the previous findings of Fisher, Franx \& Illingworth
(1995) and Thomas, Maraston \& Bender (2002) in elliptical galaxies.

Different authors have tried to explain the observed [Mg/Fe]
overabundances in the context of galaxy formation and evolution. All
these scenarios are based on nucleosynthesis models which predict that
Type II supernovae mainly eject $\alpha$-elements, like Mg, while Type
Ia supernovae are the main sources of enrichment of the ISM with
Fe. Possible explanations include variations in the initial mass
function (e.g.\ Schmidt 1963; Worthey et al.\ 1992; Vazdekis et al.\
1999; Nagashima et al.\ 2005), selective loss of metals (Worthey et
al.\ 1992), different time scales of star formation and different star
formation histories. The last two have been very popular in the last
few years in order to explain the observed overabundances in bright
ellipticals (e.g. Mehlert et al.\ 2003; S\'anchez-Bl\'azquez et al.\
2006). In this scenario, early-type galaxies are believed to be
remnants of gaseous mergers. This process produces a starburst where
part of the gas is consumed and the remainder is ejected via stellar
winds (for low mass remnants) or heated by different mechanisms (e.g.\
AGN, SNe, thermal conduction) in more massive systems, stopping
further star formation. According to the hierarchical merging
paradigm, the starburst is more intense in the central regions of the
remnant given the stronger potential well in the centre and subsequent
dissipation of the gas towards it. The merger process implies a faster
build-up of the system total mass, having shorter star formation
timescales than systems grown by secular or more gentle processes. A
short burst of star formation would not give enough time to
incorporate the gas enriched with Fe from SNe Type Ia to the formation
of further generations of stars, while SNe Type II would succeed in
incorporation $\alpha$-elements (like Mg) given their much quicker
occurrence (in less than $\rm 1\, Gyr$). In consequence, a relative
[Mg/Fe] overabundance will be observed in these systems (understood as
an \textsl{underabundance} of Fe) with respect to galaxies with
younger luminosity-weighted ages and more extended star formation
histories.

This scenario is in agreement with the results found here for the
bright S0s, and is not surprising for NGC~1316, a well known merger
remnant. However, whether this picture can be applied to the entire
NGC~1380 and NGC~1381 galaxies or only to their nuclei, is something
that cannot be tested with central line indices alone.
\subsubsection{The Residuals of the Index$^*$--log($\sigma_0$) relations}

Previous studies have looked for correlations between the
Index$^*$--log($\sigma_0$) residuals ($\rm \Delta Index^*$) and
different parameters that characterise the galaxies' structure or
their stellar populations. Some of these efforts point out the
existence of a dependence between $\rm \Delta Index^*$ and the
relative ages of elliptical galaxies (Schweizer et al.\ 1990;
Gonz\'alez \& Gorgas 1996; Worthey \& Collobert 2003) by comparing the
residuals to the index H$\beta$. Hints of similar trends are seen for
the Mg indices of these S0 galaxies but they do not seem to be
statistically significant, while residuals from the Fe relations are
clearly uncorrelated with H$\beta$. Potential correlations between
$\rm \Delta Index^*$ and other parameters were explored, including
kinematical ($\sigma_0$; maximum -circular- rotational velocity,
$V_{\rm MAX}$) and structural ($R_{\rm e}$, $R_{\rm d}$,
bulge-to-total fraction, S\'ersic index) parameters derived in
Papers~I and II. No statistically significant trends were found for
any of these combinations.

\subsection{The Index$^*$--log($V_{\rm MAX}$) relations and the Dynamical Mass}\label{sec:slpIndVmax}
\begin{figure*}
\begin{center}
\includegraphics[scale=0.72]{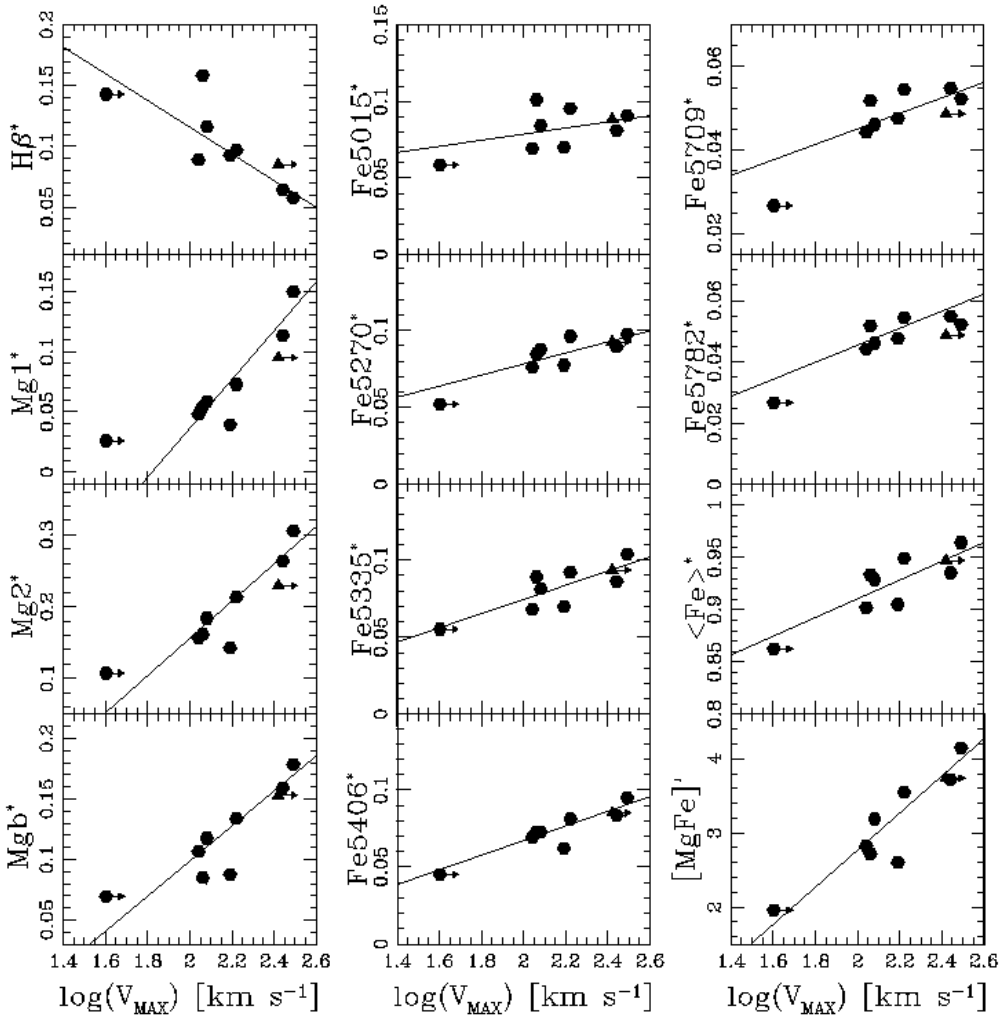}
\end{center}
\caption{\label{fig:indVmax}\small Central line indices (in magnitudes)
  versus $\log(V_{\rm MAX})$. Solid lines are the best fits to
  the datapoints, excluding ESO~359-G002. The triangle represents NGC~1316, a
  merger remnant. For the two galaxies with arrows, the deprojected azimuthal
  velocity, $V_{\phi}$, was used instead of $V_{\rm MAX}$ as a lower
  limit.}
\end{figure*}

Given the strong rotational support of many of these S0 galaxies, we
decided to explore the Index$^*$--log($V_{\rm MAX}$)
relations. Figure~\ref{fig:indVmax} shows the different line indices
versus maximum (circular) rotational velocity, where clear trends
appear in all the panels. Linear least-squares fits to the data were
performed and presented as straight lines in each panel. The galaxy
with the lowest velocity, ESO~359-G002, was excluded from the fitting
process because the slope and zero-points would be strongly affected
by this pressure-supported object for which $V_{\rm MAX}$ is highly
uncertain. The exclusion of the other pressure-supported galaxy,
NGC~1316, does not significantly affect the fitted slopes and
zero-points.

In Table~\ref{tab:indVmax} the linear fit coefficients and standard deviations,
  $\sigma_{\rm std}$, are presented. It is interesting to notice that for
  almost all the indices, $\sigma_{\rm std}$ is smaller than its counterpart
  from the Index$^*$--log($\sigma_0$) fits (see Table~\ref{tab:indsigfit}). If we
  assume that mass is the fundamental physical parameter governing the
  properties of these galaxies, the
  improvement in the fits may be interpreted in the sense that $V_{\rm MAX}$
  is a better estimator of the total dynamical mass than $\sigma_0$, which is
  not totally surprising after highlighting in Paper~I the
  predominant rotational nature of these systems. As a test, the mass of
  these galaxies was parametrised as a function of the velocity dispersion and of
  $V_{\rm MAX}$ in order to compare how central indices correlate with both
  possible descriptions. If the galaxies are supported by rotation, we can write

\begin{table}  
 \centering
 \caption[Parameters of linear fits ${\rm Index^*} = a + b\cdot \log(\rm
 V_{MAX})$ of S0 galaxies in Fornax.]{Parameters of linear fits ${\rm Index^*}
 = a + b\cdot \log(\rm V_{MAX})$ of S0 galaxies in Fornax. $\sigma_{\rm std}$ is the standard deviation about the fit. \label{tab:indVmax}} 
 \begin{tabular}{@{}lccc@{}}
  \hline
   Index  & $b\pm \delta(b)$ & $a\pm \delta(a)$ & $\sigma_{\rm std}$   \\
  \hline
  $\rm H\beta ^*$                &          $-$0.1099 $\pm$ 0.0046  &\phantom{1} 0.3354 $\pm$ 0.0108  & 0.0214 \\
  $\rm Fe5015 ^*$                &\phantom{1}  0.0197 $\pm$ 0.0030  &\phantom{1} 0.0389 $\pm$ 0.0071  & 0.0143 \\
  $\rm Mg_1 ^*$                   &\phantom{1}  0.2015 $\pm$ 0.0026  &         $-$0.3665 $\pm$ 0.0062  & 0.0233 \\
  $\rm Mg_2 ^*$                   &\phantom{1}  0.2593 $\pm$ 0.0026  &         $-$0.3633 $\pm$ 0.0061  & 0.0355 \\
  ${\rm Mg}b ^*$                 &\phantom{1}  0.1446 $\pm$ 0.0038  &         $-$0.1906 $\pm$ 0.0089  & 0.0194 \\
  $\rm Fe5270 ^*$                &\phantom{1}  0.0358 $\pm$ 0.0031  &\phantom{1} 0.0062 $\pm$ 0.0073  & 0.0070 \\
  $\rm Fe5335 ^*$                &\phantom{1}  0.0462 $\pm$ 0.0036  &         $-$0.0181 $\pm$ 0.0083  & 0.0094 \\
  $\rm Fe5406 ^*$                &\phantom{1}  0.0475 $\pm$ 0.0038  &         $-$0.0279 $\pm$ 0.0090  & 0.0057 \\
  $\rm Fe5709 ^*$                &\phantom{1}  0.0185 $\pm$ 0.0037  &\phantom{1} 0.0080 $\pm$ 0.0086  & 0.0050 \\
  $\rm Fe5782 ^*$                &\phantom{1}  0.0377 $\pm$ 0.0047  &         $-$0.0399 $\pm$ 0.0109  & 0.0098 \\
$\rm \langle Fe\rangle ^*$       &\phantom{1}  0.0895 $\pm$ 0.0053  &\phantom{1} 0.7313 $\pm$ 0.0122  & 0.0166 \\
$\rm [MgFe]'$                    &\phantom{1}  2.5019 $\pm$ 0.1147  &         $-$2.2356 $\pm$ 0.2655  & 0.2802 \\
\hline
\end{tabular}
\end{table}

\begin{equation}\label{eq:eps2}
{M} \propto {V_{\rm MAX}^2}\cdot R_{\rm d},
\end{equation}
where $M$ is the total dynamical mass and the disk scale-length, $R_{\rm d}$, was used to characterise the radius
of the galaxies because it reaches the regions where the rotation curves
become flat. If the galaxies are mainly supported by velocity dispersion, a
similar expression can be obtained,
\begin{equation}\label{eq:eps3}
{M} \propto \langle \sigma^2\rangle_{\rm R_e} \cdot R_{\rm e},
\end{equation}
where $R_{\rm e}$ is the effective radius of the bulge and $\langle
\sigma^2\rangle_{\rm R_e}$ is the average mean-square dispersion
inside $R_{\rm e}$.  Other measures of scale such as the half-light
radius determined in Paper 2 produce similar results, but for this
analysis $R_e$ is adopted because it is well constrained by the {\tt
GIM2D} models (Simard et al.\ 2002).

\begin{figure*}
\begin{center}
\includegraphics[scale=0.72]{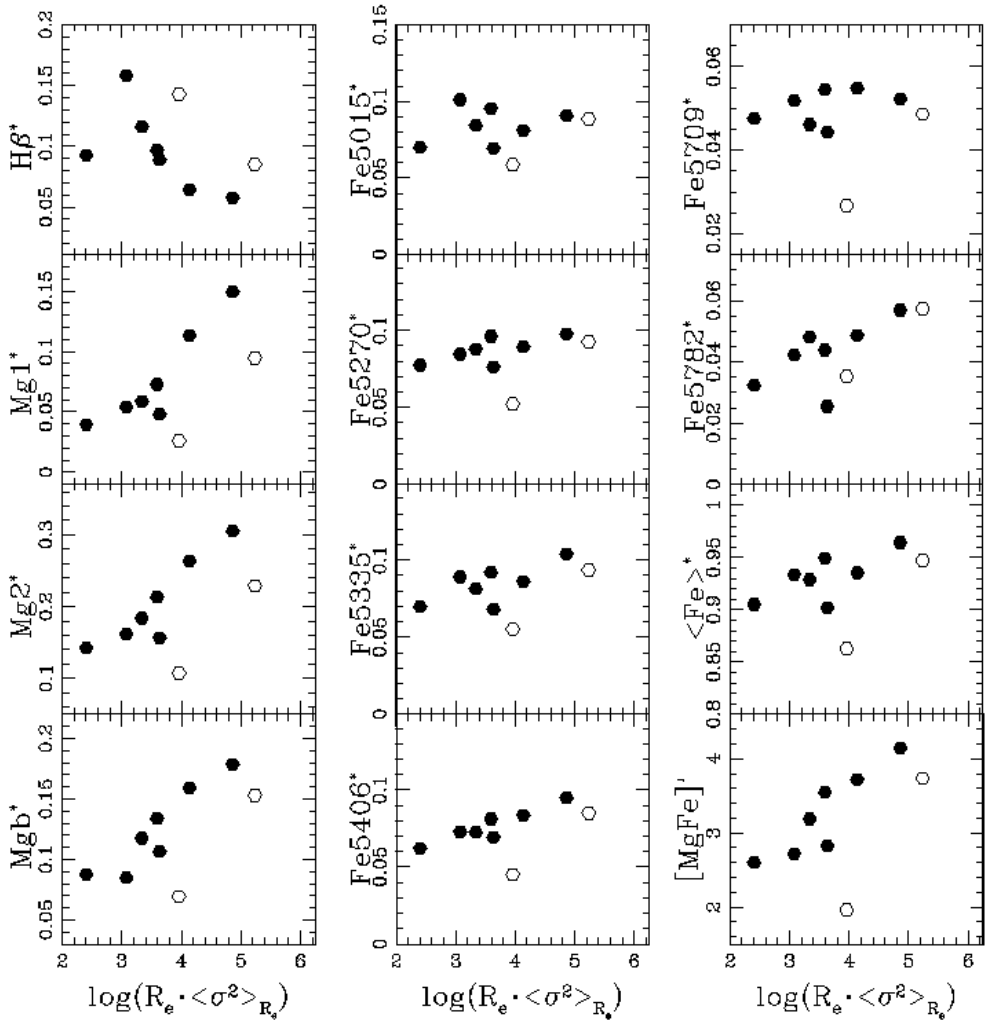}
\end{center}
\caption[Central line indices versus $\log(R_{\rm e} \cdot
  \sigma_{\rm \langle R_e \rangle}^2)$.]{\label{fig:indMasS}\small Central
  line indices versus $\log(R_{\rm e} \cdot \sigma_{\rm \langle R_e \rangle}^2)$, proportional to the total dynamical mass for pressure-supported galaxies. Open symbols represent NGC~1316 and ESO~359-G002, two pressure supported galaxies.}
\end{figure*}

\begin{figure*}
\begin{center}
\includegraphics[scale=0.72]{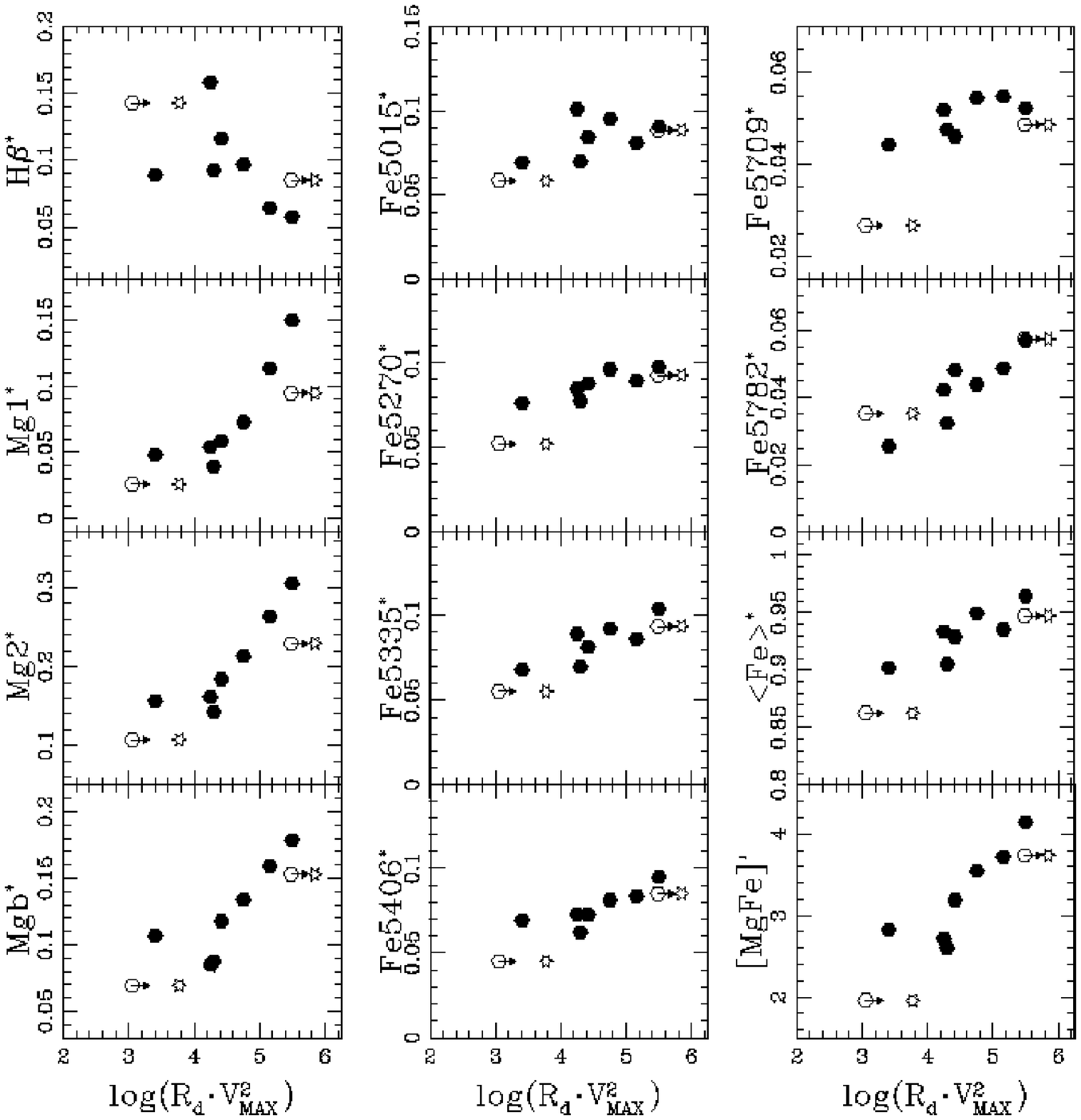}
\end{center}
\caption[Central line indices versus $\log(R_{\rm d} \cdot V_{\rm
    MAX}^2)$.]{\label{fig:indMasV}\small Central line indices versus $\log(R_{\rm d} \cdot V_{\rm MAX}^2)$, proportional to the total dynamical
  mass for rotationally-supported galaxies. Open symbols with arrows represent
  NGC~1316 and ESO~359-G002, two pressure supported galaxies. For these two
  objects the deprojected azimuthal velocity, $V_{\phi}$, was used as a lower
  limit instead of $V_{\rm MAX}$. Open stars correspond to the dynamical mass
  of these two galaxies estimated from $\sigma$ by assuming an isothermal
  sphere in hydrostatic equilibrium.}

\end{figure*}

In Figures~\ref{fig:indMasS} and \ref{fig:indMasV}, central line indices are
plotted versus $R_{\rm e}\cdot \rm \langle \sigma^2\rangle_{R_e}$ and $R_{\rm
  d}\cdot V_{\rm MAX}^2$, respectively. Clearly, both
parameters, $\sigma$ and $V_{\rm MAX}$, are tracing, to some extent, the mass of these
systems. However, an important difference arises between pressure- and
rotation-supported galaxies (open and filled dots, respectively): when the
mass is estimated using $\sigma$, it is underestimated in rotationally
supported systems with respect to pressure-supported ones, as it is evident
from the offset between open and filled dots in Figure~\ref{fig:indMasS}. On the
other hand, Figure~\ref{fig:indMasV} shows that if we estimate the mass using
$V_{\rm MAX}$, all the galaxies follow a common sequence. In this plot, lower
limits for the two pressure-supported S0s (using $V_{\phi}$) are indicated as
open dots with arrows. For these two galaxies, masses were also estimated from
their measured $\sigma$'s by assuming a simple isothermal sphere in
hydrostatic equilibrium and presented as open stars in
Figure~\ref{fig:indMasV}. The much tighter trends observed in this plot are
consistent with our hypothesis that $V_{\rm MAX}$ is a much better tracer of the
dynamical mass of these systems than $\sigma$.


\begin{figure*}
\begin{center}
\includegraphics[scale=0.75]{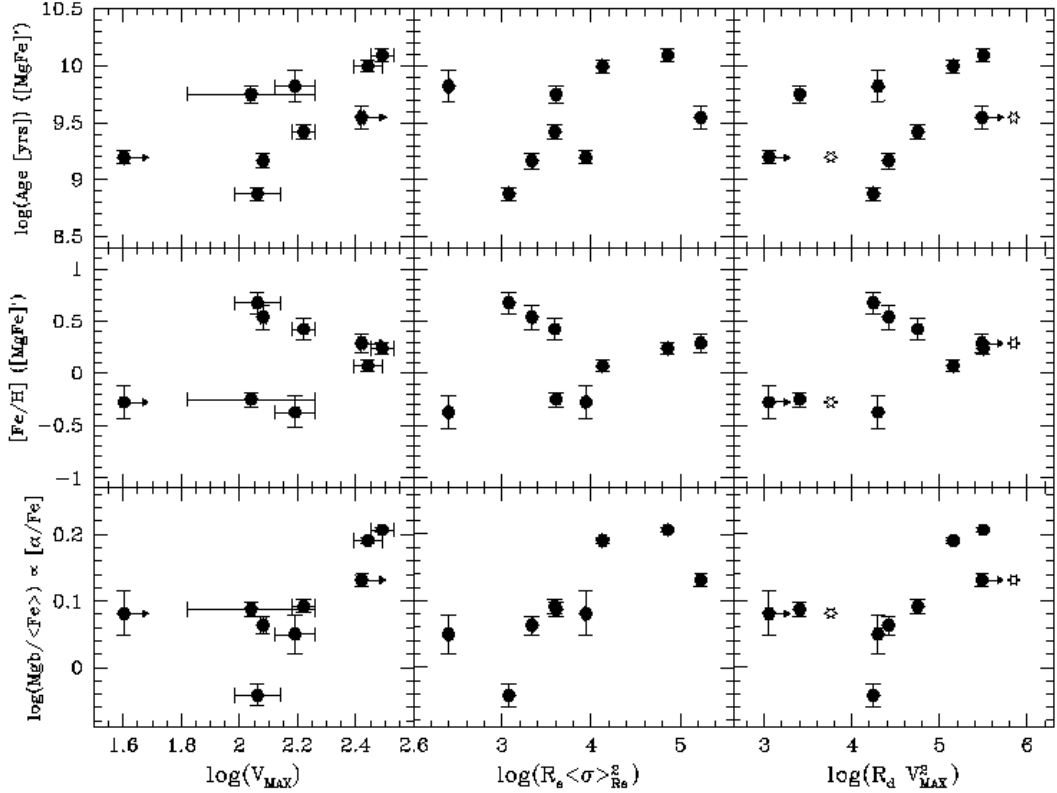}
\end{center}
\caption[Central ages, metallicities, Mg$b$/$\rm \langle Fe \rangle$
  ($\propto \alpha$-elements overabundance) versus 3 different dynamical
  mass tracers.]{\label{fig:AgeZAbundvs3Mass}\small Central ages
  and metallicities from BC03 models and Mg$b$/$\rm \langle Fe \rangle$ ($\propto \alpha$-elements
  overabundance) versus 3 different dynamical
  mass tracers. The same symbols are used as in Figure~\ref{fig:indMasV}.}
\end{figure*}

\begin{figure*}
\begin{center}
\includegraphics[scale=0.75]{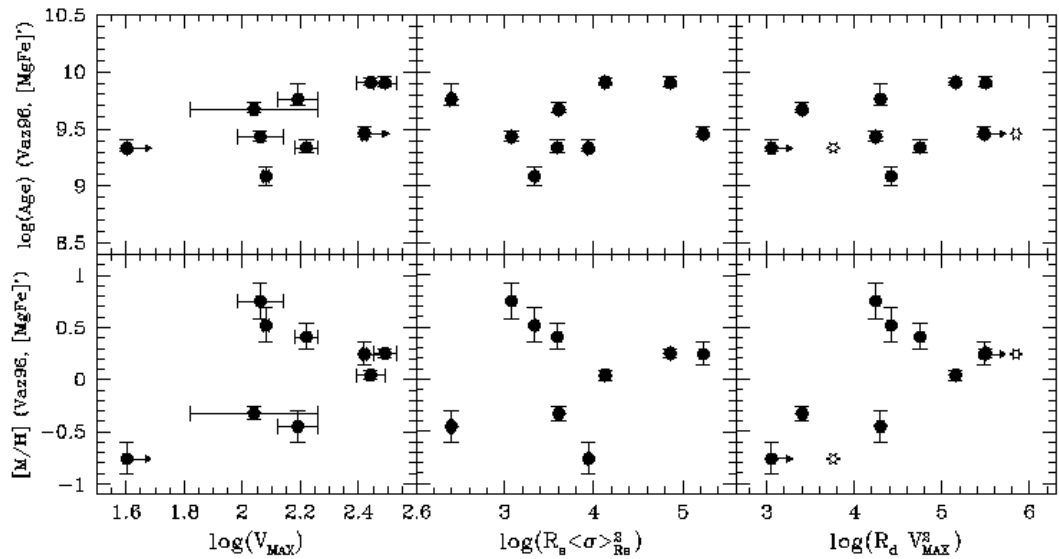}
\end{center}
\caption[Central ages and metallicities from Vaz96 models versus 3 different
dynamical mass tracers.]{\label{fig:AgeZVaz96vs3Mass}\small Central ages and
  metallicities from an improved version of Vaz96 models (Vazdekis et al.\ in
  preparation) versus 3 different dynamical mass tracers. The same symbols are used as in Figure~\ref{fig:indMasV}.}
\end{figure*}

Figure~\ref{fig:AgeZAbundvs3Mass} presents central ages and
[Fe/H] from BC03 models and Mg$b$/$\rm \langle Fe\rangle$ versus 3 different dynamical mass
tracers ($V_{\rm MAX}$, $R_{\rm e}\cdot \rm \langle \sigma^2\rangle_{R_e}$ and $R_{\rm d}\cdot V_{\rm MAX}^2$). No strong correlations
are obvious from the central stellar population properties plotted
against the dynamical mass. However, for $R_{\rm d}\cdot V_{\rm MAX}^2$ mass
tracer, a Spearman rank and
a Student-t correlation tests (Table~\ref{tab:physCorrTests}) indicate
that the relation between the [$\alpha$/Fe] tracer and
dynamical mass is significant at a 95--99\% confidence level. To understand
the absence of other correlations, we explore the possibility that the
uncertainties in BC03 models are washing out the real trends with age and
metallicity by considering other SSP models. We have used an improved version of Vazdekis et al.\
(1996) models (Vazdekis et al.\, in preparation; hereafter Vaz96) and
recalculate all central ages and metallicities ([M/H] in Vaz96 nomenclature)
for our S0 sample \footnote{We have used H$\rm \beta$ and [MgFe]' indices at
  Lick resolution as our age and metallicity indicators, respectively.}. In
Figure~\ref{fig:AgeZVaz96vs3Mass}, we present the resulting ages and [M/H]
versus different mass tracers. No strong correlations are observed either with
the new model predictions, weakening the assumption that model artifacts are
responsible of the absence of trends between the stellar population parameters
and the dynamical mass. An alternative explanation is that the tight
correlations observed between the line indices and galaxy mass are not
governed by the variation of one of these stellar population parameters alone,
but by a combination of them in some correlated way.

\begin{figure}
\begin{center}
\includegraphics[scale=0.65]{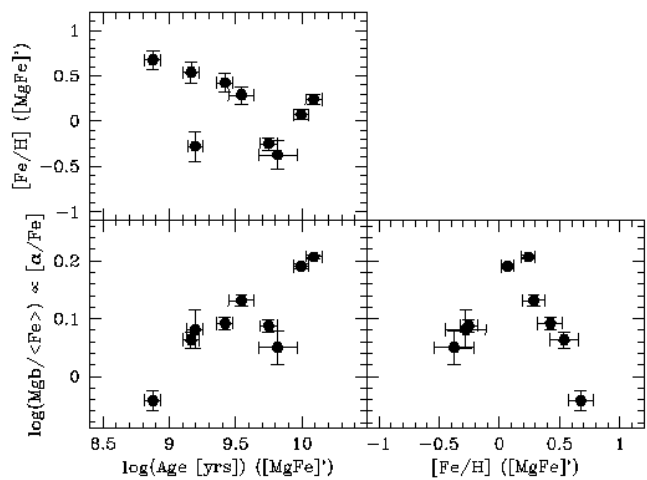}
\end{center}
\caption[Central ages, metallicities, Mg$b$/$\rm \langle Fe \rangle$
  ($\propto \alpha$-elements overabundance) versus
  eachother.]{\label{fig:AgeZAbundvseachother}\small Central ages and 
  metallicities from BC03 models and Mg$b$/$\rm \langle Fe \rangle$ ($\propto \alpha$-elements
  overabundance) versus eachother.}
\end{figure}

Indeed, Figure~\ref{fig:AgeZAbundvseachother} suggests that age and
$\alpha$-element overabundance are correlated  (97.5\% confidence level). High
$\alpha$-element overabundance implies short star-formation timescale. This can
be understood by considering a simple model for the star-formation history of
the Fornax S0s in which all these galaxies started forming stars at
approximately the same time and formed stars for a period $\Delta t$.
Thereafter, star formation ceased and the galaxies evolved passively for a
time $t$ until the epoch of observation. The age derived from the
absorption-line indices is $\simeq t$, while the  $\alpha$-element
overabundance is anticorrelated with $\Delta t$. In this simple model $t$ is
anticorrelated with $\Delta t$, and it follows that the measured age must be
correlated with the $\alpha$-element overabundance, as observed.  If this
interpretation is correct, the most massive S0s have the shortest
star-formation timescales (cf. \S5.1.2) and the oldest stellar populations.
Obviously, this is an oversimplified model, but it explains, at lest
qualitatively, the observed trends. 

\begin{table}  
 \centering
 \caption[Spearman and Student-t tests for central stellar population
 parameters versus dynamical mass of S0 galaxies in Fornax.]{ Spearman and Student-t tests for central stellar population
 parameters versus dynamical mass of S0 galaxies in Fornax
 (as estimated from $R_{\rm d}\cdot V_{\rm MAX}^2$, third column in Figure~\ref{fig:AgeZAbundvs3Mass}). In parenthesis, probability of rejecting the null
 hypothesis of no correlation.\label{tab:physCorrTests}} 
 \begin{tabular}{@{}lll@{}}
  \hline
   Dyn. Mass vs.  & Spearman Test Coeff. & Student-t test Coeff. \\
  \hline
  $\rm \log({age})_{[MgFe]'}$                       & 0.4167 ($<$ 90\%)      & 1.2127 (85--90\%) \\
  $\rm [Fe/H]_{[MgFe]'}$               & 0.2833 ($<$ 90\%)      & 0.7817 (75--80\%) \\
  Mg$b$/$\rm \langle Fe \rangle$ & 0.6833 (95--97.5\%)  & 2.4762 (97.5--99\%) \\
\hline
\end{tabular}
\end{table}

\section{Conclusions}\label{sec:conc}

This paper presents the study of a sample of 9 S0 galaxies in the  Fornax
Cluster previously described in Paper~I. By linking the properties of the
central stellar populations (ages, chemical abundances) of these galaxies  and
their global properties (mass and dynamics), we uncover important clues on the
main physical drivers governing the formation and evolution of S0s.  Our main
conclusions are the following:

\begin{itemize}

\item Central absorption line indices correlate with central velocity
  dispersions in a way similar to what previous studies found for elliptical
  galaxies. However, a study of the stellar population properties of the
  Fornax S0s reveals that the trends shown by their line indices seem to be
  produced by relative differences in age and $\alpha$-element abundances,
  contrary to what is found in ellipticals where the overall metallicities are
  important drivers of the correlations.

\item The scatter in the Index$^*$--$\log(\sigma_0)$ relations can be
  partially explained by the rotationally-supported nature of many
  S0s. The tighter correlations found between Index$^*$ and
  $\log(V_{\rm MAX})$ support this interpretation.

\item The dynamical mass seems to be the primary physical property
  governing these correlations and in the Fornax S0s we need to study
  it by considering their rotationally-supported nature. For these
  systems, $V_{\rm MAX}$ is a better tracer of dynamical mass than
  $\sigma$.
  
\item The $\alpha$-element overabundance of these galaxies seems to be
  correlated with their mass. Moreover, 
  the absorption-line-derived ages also 
  correlate with the overabundances. This implies that the most massive 
  S0s have the shortest star-formation timescales and the oldest stellar
  populations.

\end{itemize}
 
These results support the idea of different star formation histories
for the central regions of bright and faint S0 galaxies. Other authors
have reached similar conclusions by studying early-type systems in the
Coma Cluster (Poggianti et al.\ 2001; Mehlert et al.\ 2003).

The results found here are consistent with a scenario where faint S0s
are descendants of spiral galaxies who lost (or exhausted) their gas
and suffered a final central gasp of star formation during cluster
infall, while bright systems seem to have star formation
histories resembling those of normal ellipticals. 
However, are the nuclei characterising the entire
galaxies? Have the outermost regions another story to tell? These
questions lie beyond the scope of this study, but will be addressed in
the following paper of the series, when a stellar population study at
larger galactocentric distances will give us new clues about the
detailed star formation history of S0s in the Fornax Cluster.

\section*{Acknowledgments}

This work was based on observations made with ESO telescopes at Paranal
Observatory under programme ID 070.A-0332. This publication makes use of data
products from the Two Micron All Sky Survey, which is a joint project of the
University of Massachusetts and the Infrared Processing and Analysis
Center/California Institute of Technology, funded by the National
Aeronautics and Space Administration and the National Science
Foundation.

\appendix

\section{Tables}

In this appendix we include tables with different parameters for each galaxy
of our sample.\\

Table~A1 includes central absorption line indices measured at $3\,\rm \AA$
resolution for all our galaxies while Table~A2 presents similar values
measured at Lick resolution.

In Table~A3, central luminosity-weighted ages and [Fe/H] are presented for our
galaxies by using different [Fe/H] indicators, H$\beta$ as the age indicator
and Bruzual \& Charlot (2003) models at Lick resolution.

\begin{landscape}
\begin{table}
\begin{center}
\caption{\label{tab:indcent3A}Central Line Indices measured at $\rm 3\, \AA$ resolution of S0 galaxies in Fornax.}
\begin{tabular}{@{}lc@{\,\,\,}c@{\,\,\,}c@{\,\,\,}c@{\,\,\,}c@{\,\,\,}c@{\,\,\,}c@{\,\,\,}c@{\,\,\,}c@{\,\,\,}c@{\,\,\,}c@{\,\,\,}c@{}}
\hline
\hline
 Name & $\rm [MgFe]'$ & $\rm \langle Fe \rangle$ & H$\beta$ & Fe5015 & $\rm Mg_1$ &
     $\rm Mg_2$ & Mg$b$ & Fe5270 & Fe5335 & Fe5406 & Fe5709 &  Fe5782 \\
   &  [\AA] & [\AA] &[\AA] & [\AA] & [mag] & [mag] &[\AA] & [\AA] & [\AA] & [\AA] & [\AA] & [\AA] \\

    (1)   &      (2) &     (3)  &   (4)  &    (5)    &     (6)    &   (7)   &
          (8) & (9) & (10) & (11) & (12) & (13)  \\
\hline
\hline
CENTRAL\\
\hline\\[0.2em]
                   NGC\,1380& 4.14\,(0.04)& 3.53\,(0.03) & 1.48\,(0.04) & 6.10\,(0.08)& 0.149\,(0.001)& 0.305\,(0.001)& 4.92\,(0.03)& 3.43\,(0.04) & 3.64\,(0.04) & 2.29\,(0.03)& 1.11\,(0.03)& 1.02\,(0.03)\\\\
                   NGC\,1381& 3.71\,(0.04)& 3.09\,(0.03) & 1.65\,(0.04) & 5.46\,(0.08)& 0.113\,(0.001)& 0.263\,(0.001)& 4.42\,(0.03)& 3.14\,(0.04) & 3.04\,(0.04) & 2.03\,(0.03)& 1.16\,(0.03)& 0.87\,(0.03)\\\\
                  NGC\,1380A& 3.18\,(0.09)& 2.99\,(0.06) & 2.91\,(0.08) & 5.68\,(0.16)& 0.058\,(0.001)& 0.183\,(0.002)& 3.33\,(0.08)& 3.10\,(0.08) & 2.88\,(0.10) & 1.77\,(0.07)& 0.98\,(0.07)& 0.86\,(0.07)\\\\
                   NGC\,1375& 2.71\,(0.08)& 3.06\,(0.06) & 3.89\,(0.09) & 6.76\,(0.17)& 0.053\,(0.001)& 0.161\,(0.002)& 2.43\,(0.08)& 2.98\,(0.08) & 3.15\,(0.09) & 1.78\,(0.06)& 1.10\,(0.05)& 0.76\,(0.05)\\\\
                    IC\,1963& 3.54\,(0.08)& 3.31\,(0.05) & 2.45\,(0.07) & 6.40\,(0.14)& 0.072\,(0.001)& 0.212\,(0.001)& 3.76\,(0.07)& 3.38\,(0.07) & 3.24\,(0.08) & 1.97\,(0.06)& 1.16\,(0.05)& 0.79\,(0.05)\\\\
\scriptsize{ESO\,358$-$G006}& 2.59\,(0.16)& 2.61\,(0.11) & 2.35\,(0.15) & 4.75\,(0.29)& 0.039\,(0.003)& 0.142\,(0.003)& 2.51\,(0.14)& 2.75\,(0.15) & 2.48\,(0.17) & 1.52\,(0.13)& 1.01\,(0.11)& 0.58\,(0.12)\\\\
\scriptsize{ESO\,358$-$G059}& 2.82\,(0.07)& 2.56\,(0.05) & 2.25\,(0.06) & 4.71\,(0.12)& 0.048\,(0.001)& 0.155\,(0.001)& 3.03\,(0.06)& 2.70\,(0.06) & 2.42\,(0.07) & 1.69\,(0.05)& 0.94\,(0.05)& 0.46\,(0.05)\\\\
                   NGC\,1316& 3.73\,(0.08)& 3.28\,(0.05) & 2.16\,(0.07) & 5.96\,(0.14)& 0.094\,(0.014)& 0.228\,(0.001)& 4.26\,(0.07)& 3.26\,(0.07) & 3.30\,(0.08) & 2.07\,(0.06)& 1.04\,(0.05)& 1.02\,(0.05)\\\\
\scriptsize{ESO\,359$-$G002}& 1.95\,(0.15)& 1.92\,(0.11) & 3.54\,(0.13) & 3.99\,(0.26)& 0.025\,(0.011)& 0.107\,(0.003)& 2.01\,(0.13)& 1.87\,(0.14) & 1.97\,(0.16) & 1.12\,(0.12)& 0.57\,(0.11)& 0.63\,(0.11)\\[0.5em]
\hline
\end{tabular}\\
\end{center}
\footnotesize{Notes: From (2) to (13), $1\,\sigma$ rms errors
  between $\rm ^{''}(\,)^{''}$; Col (1), name; Col (2), $\rm [MgFe]^{\prime}$ combined index (Gonz\'alez
  1993; Thomas, Maraston \& Bender 2003); Col (3), $\rm \langle Fe \rangle$ combined index (Gorgas, Efstathiou \&
  Arag\'on-Salamanca 1990); Col (4), H$\beta$ index; Col (5), Fe5015 index;
  Col (6), $\rm Mg_1$ index in magnitudes; Col (7), $\rm Mg_2$ index in magnitudes; Col (8),
  Mg$b$ index; Col (9), Fe5270 index; Col (10), Fe5335 index; Col (11), Fe5406
  index; Col (12), Fe5709 index; Col (13), Fe5782 index.}
\end{table}
\end{landscape}

\begin{landscape}
\begin{table}
\begin{center}
\caption{\label{tab:indcentLick}Central Line Indices measured at Lick resolution of S0 galaxies in Fornax.}
\begin{tabular}{@{}lc@{\,\,\,}c@{\,\,\,}c@{\,\,\,}c@{\,\,\,}c@{\,\,\,}c@{\,\,\,}c@{\,\,\,}c@{\,\,\,}c@{\,\,\,}c@{\,\,\,}c@{\,\,\,}c@{}}
\hline
\hline
 Name & $\rm [MgFe]'$ & $\rm \langle Fe \rangle$ & H$\beta$ & Fe5015 & $\rm Mg_1$ &
     $\rm Mg_2$ & Mg$b$ & Fe5270 & Fe5335 & Fe5406 & Fe5709 &  Fe5782 \\
   &  [\AA] & [\AA] &[\AA] & [\AA] & [mag] & [mag] &[\AA] & [\AA] & [\AA] & [\AA] & [\AA] & [\AA] \\

    (1)   &      (2) &     (3)  &   (4)  &    (5)    &     (6)    &   (7)   &
          (8) & (9) & (10) & (11) & (12) & (13)  \\
\hline
\hline
CENTRAL \\
\hline\\[0.2em]
                   NGC\,1380& 3.82\,(0.04)& 3.01\,(0.03) & 1.47\,(0.04) & 5.13\,(0.08)& 0.162$^{a}$\,(0.007)& 0.322$^{a}$\,(0.007)& 4.86\,(0.04)& 3.02\,(0.04) & 3.00\,(0.04) & 1.97\,(0.03)& 1.00\,(0.03)& 0.83\,(0.03)\\\\
                   NGC\,1381& 3.40\,(0.04)& 2.70\,(0.03) & 1.65\,(0.04) & 4.83\,(0.08)& 0.128$^{a}$\,(0.007)& 0.283$^{a}$\,(0.007)& 4.20\,(0.04)& 2.81\,(0.04) & 2.60\,(0.05) & 1.75\,(0.03)& 1.05\,(0.03)& 0.75\,(0.03)\\\\
                  NGC\,1380A& 2.88\,(0.09)& 2.65\,(0.06) & 2.78\,(0.08) & 4.69\,(0.16)& 0.075$^{a}$\,(0.011)& 0.204$^{a}$\,(0.007)& 3.07\,(0.08)& 2.79\,(0.09) & 2.51\,(0.10) & 1.53\,(0.07)& 0.90\,(0.07)& 0.66\,(0.07)\\\\
                   NGC\,1375& 2.52\,(0.07)& 2.63\,(0.05) & 3.47\,(0.07) & 5.24\,(0.14)& 0.057$^{a}$\,(0.011)& 0.169$^{a}$\,(0.007)& 2.39\,(0.07)& 2.69\,(0.07) & 2.57\,(0.08) & 1.55\,(0.00.1626)& 1.00\,(0.05)& 0.68\,(0.05)\\\\
                    IC\,1963& 3.22\,(0.08)& 2.87\,(0.05) & 2.26\,(0.07) & 5.27\,(0.14)& 0.089$^{a}$\,(0.015)& 0.233$^{a}$\,(0.007)& 3.56\,(0.07)& 3.00\,(0.07) & 2.75\,(0.08) & 1.71\,(0.06)& 1.02\,(0.05)& 0.64\,(0.05)\\\\
\scriptsize{ESO\,358$-$G006}& 2.41\,(0.16)& 2.23\,(0.12) & 2.11\,(0.15) & 3.97\,(0.29)& 0.054$^{a}$\,(0.028)& 0.161$^{a}$\,(0.008)& 2.51\,(0.14)& 2.44\,(0.15) & 2.03\,(0.18) & 1.35\,(0.13)& 0.89\,(0.12)& 0.50\,(0.12)\\\\
\scriptsize{ESO\,358$-$G059}& 2.53\,(0.07)& 2.25\,(0.05) & 2.16\,(0.06) & 3.92\,(0.13)& 0.065$^{a}$\,(0.008)& 0.178$^{a}$\,(0.007)& 2.76\,(0.06)& 2.40\,(0.06) & 2.10\,(0.07) & 1.47\,(0.05)& 0.87\,(0.05)& 0.34\,(0.05)\\\\
                   NGC\,1316& 3.22\,(0.08)& 2.76\,(0.05) & 2.11\,(0.07) & 5.15\,(0.14)& 0.111$^{a}$\,(0.015)& 0.250$^{a}$\,(0.007)& 3.74\,(0.07)& 2.77\,(0.07) & 2.75\,(0.08) & 1.66\,(0.06)& 0.93\,(0.05)& 0.77\,(0.05)\\\\
\scriptsize{ESO\,359$-$G002}& 1.70\,(0.14)& 1.54\,(0.10) & 3.32\,(0.13) & 3.15\,(0.25)& 0.039$^{a}$\,(0.012)& 0.123$^{a}$\,(0.008)& 1.86\,(0.13)& 1.58\,(0.14) & 1.51\,(0.16) & 1.00\,(0.12)& 0.52\,(0.11)& 0.51\,(0.11)\\[0.5em]
\hline
\end{tabular}\\
\end{center}
\footnotesize{Notes: From (2) to (13), $1\,\sigma$ rms errors
  between $\rm ^{''}(\,)^{''}$; Col (1), name; Col (2), $\rm [MgFe]^{'}$ combined index (Gonz\'alez
  1993; Thomas, Maraston \& Bender 2003); Col (3), $\rm \langle Fe \rangle$ combined index (Gorgas, Efstathiou \&
  Arag\'on-Salamanca 1990); Col (4), H$\beta$ index; Col (5), Fe5015 index;
  Col (6), $\rm Mg_1$ index in magnitudes; Col (7), $\rm Mg_2$ index in magnitudes; Col (8),
  Mg$b$ index; Col (9), Fe5270 index; Col (10), Fe5335 index; Col (11), Fe5406
  index; Col (12), Fe5709 index; Col (13), Fe5782 index.\\ ($^{a}$) An
  additive spectrophotometric correction to the Lick system was applied to the
  indices $\rm Mg_1$ and $\rm Mg_2$.}
\end{table}
\end{landscape}

\begin{landscape}
\begin{table}
\begin{center}
\caption{\label{tab:agezcent}Central Ages and Metallicities using line indices at Lick resolution for S0 galaxies in Fornax.}
\begin{tabular}{@{}lc@{\,\,\,}c@{\,\,\,}c@{\,\,\,}c@{\,\,\,}c@{\,\,\,}c@{\,\,\,}c@{\,\,\,}c@{}}
\hline
\hline
 Name & $\rm \log(Age)_{[MgFe]'}$ & $\rm [Fe/H]_{[MgFe]'}$ & $\rm \log(Age)_{\langle Fe \rangle}$ & $\rm [Fe/H]_{\langle Fe \rangle}$ & $\rm \log(Age)_{Mgb}$ & $\rm [Fe/H]_{Mgb}$ & $\rm \log(Age)_{Fe5709}$ & $\rm [Fe/H]_{Fe5709}$  \\
     & [yr] &  [dex]  & [yr] & [dex] & [yr] & [dex] & [yr] & [dex]\\

    (1)   &      (2) &     (3)  &   (4)  &    (5)    &     (6)    &   (7)   &  (8) & (9)   \\
\hline
\hline\\[0.2em]
                   NGC\,1380&  $ 10.09_{\phantom{1}-0.05}^{\phantom{1}+0.06}$ &  $\phantom{1}  0.23_{\phantom{1}-0.05}^{\phantom{1}+0.05}$ & $ 10.21_{\phantom{1}-0.06}^{\phantom{1}+0.06}$ & $ -0.02_{\phantom{1}-0.03}^{\phantom{1}+0.03}$ & $ 9.95_{\phantom{1}-0.05}^{\phantom{1}+0.05}$ & $\phantom{1}  0.63_{\phantom{1}-0.06}^{\phantom{1}+0.06}$ & $ 10.23_{\phantom{1}-0.05}^{\phantom{1}+0.06}$ & $ -0.05_{\phantom{1}-0.05}^{\phantom{1}+0.05}$ \\\\
                   NGC\,1381&  $  9.99_{\phantom{1}-0.05}^{\phantom{1}+0.05}$ &  $\phantom{1}  0.06_{\phantom{1}-0.05}^{\phantom{1}+0.05}$ & $ 10.10_{\phantom{1}-0.05}^{\phantom{1}+0.06}$ & $ -0.17_{\phantom{1}-0.04}^{\phantom{1}+0.03}$ & $ 9.89_{\phantom{1}-0.06}^{\phantom{1}+0.06}$ & $\phantom{1}  0.34_{\phantom{1}-0.06}^{\phantom{1}+0.07}$ & $  9.99_{\phantom{1}-0.04}^{\phantom{1}+0.04}$ & $\phantom{1}  0.06_{\phantom{1}-0.06}^{\phantom{1}+0.06}$ \\\\
                  NGC\,1380A&  $  9.16_{\phantom{1}-0.06}^{\phantom{1}+0.06}$ &  $\phantom{1}  0.53_{\phantom{1}-0.11}^{\phantom{1}+0.12}$ & $  9.23_{\phantom{1}-0.04}^{\phantom{1}+0.04}$ & $\phantom{1}  0.33_{\phantom{1}-0.06}^{\phantom{1}+0.06}$ & $ 9.08_{\phantom{1}-0.08}^{\phantom{1}+0.08}$ & $\phantom{1}  0.77_{\phantom{1}-0.15}^{\phantom{1}+0.16}$ & $  9.30_{\phantom{1}-0.04}^{\phantom{1}+0.04}$ & $\phantom{1}  0.07_{\phantom{1}-0.17}^{\phantom{1}+0.15}$ \\\\
                   NGC\,1375&  $  8.87_{\phantom{1}-0.05}^{\phantom{1}+0.05}$ &  $\phantom{1}  0.67_{\phantom{1}-0.10}^{\phantom{1}+0.10}$ & $  8.91_{\phantom{1}-0.04}^{\phantom{1}+0.04}$ & $\phantom{1}  0.56_{\phantom{1}-0.04}^{\phantom{1}+0.04}$ & $ 8.84_{\phantom{1}-0.07}^{\phantom{1}+0.06}$ & $\phantom{1}  0.78_{\phantom{1}-0.13}^{\phantom{1}+0.14}$ & $  8.97_{\phantom{1}-0.05}^{\phantom{1}+0.04}$ & $\phantom{1}  0.36_{\phantom{1}-0.09}^{\phantom{1}+0.08}$ \\\\
                    IC\,1963&  $  9.41_{\phantom{1}-0.06}^{\phantom{1}+0.05}$ &  $\phantom{1}  0.42_{\phantom{1}-0.10}^{\phantom{1}+0.10}$ & $  9.43_{\phantom{1}-0.05}^{\phantom{1}+0.06}$ & $\phantom{1}  0.27_{\phantom{1}-0.05}^{\phantom{1}+0.05}$ & $ 9.36_{\phantom{1}-0.07}^{\phantom{1}+0.07}$ & $\phantom{1}  0.59_{\phantom{1}-0.13}^{\phantom{1}+0.14}$ & $  9.47_{\phantom{1}-0.06}^{\phantom{1}+0.07}$ & $\phantom{1}  0.14_{\phantom{1}-0.11}^{\phantom{1}+0.11}$ \\\\
\scriptsize{ESO\,358$-$G006}&  $  9.81_{\phantom{1}-0.13}^{\phantom{1}+0.14}$ &  $ -0.37_{\phantom{1}-0.15}^{\phantom{1}+0.16}$ & $  9.79_{\phantom{1}-0.12}^{\phantom{1}+0.13}$ & $ -0.30_{\phantom{1}-0.12}^{\phantom{1}+0.12}$ & $ 9.83_{\phantom{1}-0.13}^{\phantom{1}+0.14}$ & $ -0.43_{\phantom{1}-0.13}^{\phantom{1}+0.14}$ & $  9.72_{\phantom{1}-0.14}^{\phantom{1}+0.15}$ & $ -0.10_{\phantom{1}-0.30}^{\phantom{1}+0.25}$ \\\\
\scriptsize{ESO\,358$-$G059}&  $  9.74_{\phantom{1}-0.07}^{\phantom{1}+0.06}$ &  $ -0.25_{\phantom{1}-0.07}^{\phantom{1}+0.07}$ & $  9.75_{\phantom{1}-0.06}^{\phantom{1}+0.06}$ & $ -0.26_{\phantom{1}-0.05}^{\phantom{1}+0.05}$ & $ 9.74_{\phantom{1}-0.08}^{\phantom{1}+0.07}$ & $ -0.23_{\phantom{1}-0.06}^{\phantom{1}+0.07}$ & $  9.70_{\phantom{1}-0.06}^{\phantom{1}+0.06}$ & $ -0.13_{\phantom{1}-0.12}^{\phantom{1}+0.11}$ \\\\
                   NGC\,1316&  $  9.54_{\phantom{1}-0.10}^{\phantom{1}+0.09}$ &  $\phantom{1}  0.28_{\phantom{1}-0.09}^{\phantom{1}+0.09}$ & $  9.61_{\phantom{1}-0.08}^{\phantom{1}+0.07}$ & $\phantom{1}  0.12_{\phantom{1}-0.05}^{\phantom{1}+0.05}$ & $ 9.41_{\phantom{1}-0.07}^{\phantom{1}+0.09}$ & $\phantom{1}  0.57_{\phantom{1}-0.13}^{\phantom{1}+0.13}$ & $  9.69_{\phantom{1}-0.08}^{\phantom{1}+0.08}$ & $ -0.04_{\phantom{1}-0.12}^{\phantom{1}+0.11}$ \\\\
\scriptsize{ESO\,359$-$G002}&  $  9.19_{\phantom{1}-0.05}^{\phantom{1}+0.05}$ &  $ -0.27_{\phantom{1}-0.15}^{\phantom{1}+0.16}$ & $  9.23_{\phantom{1}-0.05}^{\phantom{1}+0.05}$ & $ -0.45_{\phantom{1}-0.10}^{\phantom{1}+0.10}$ & $ 9.13_{\phantom{1}-0.07}^{\phantom{1}+0.06}$ & $ -0.01_{\phantom{1}-0.15}^{\phantom{1}+0.17}$ & $  9.37_{\phantom{1}-0.13}^{\phantom{1}+0.17}$ & $ -0.89_{\phantom{1}-0.42}^{\phantom{1}+0.37}$ \\[0.5em]
\hline
\end{tabular}\\
\end{center}
\footnotesize{Notes: From (2) to (9), $99\%$ confidence intervals are
  presented. Col (1), name; Col (2) (3), age and metallicity, estimated using Bruzual \& Charlot
  (2003) simple stellar population models at Lick resolution and $\rm [MgFe]'$ as metallicity indicator;
  Col (4) (5), age and metallicity, estimated using $\rm \langle Fe \rangle$ as
  metallicity indicator; Col (6) (7), age and metallicity, estimated using Mg$b$ as metallicity indicator; Col (8) (9), age and metallicity, estimated using Fe5709 as metallicity indicator.
  }
\end{table}
\end{landscape}

\bsp

\label{lastpage}

\end{document}